\documentclass[preprint,12pt]{article}

\usepackage{amssymb}
\usepackage{amsmath}
\usepackage{graphicx}
\usepackage{multirow}
\usepackage{amsmath,amssymb,amsfonts}
\usepackage{amsthm}
\usepackage{mathtools}
\usepackage{caption}
\usepackage{mathrsfs}
\usepackage[title]{appendix}
\usepackage{url}
\usepackage{xcolor}
\usepackage{colortbl}
\usepackage{textcomp}
\usepackage{manyfoot}
\usepackage{booktabs}
\usepackage{algorithm}
\usepackage{algorithmicx}
\usepackage{algpseudocode}
\usepackage{listings}
\usepackage{pifont}
\definecolor{mistyrose}{rgb}{1.0, 0.89, 0.88}
\usepackage{a4wide}
\usepackage{enumitem}

\newcommand{\net}{DNNCS\ }
\newcommand{\cmark}{\ding{51}}%
\newcommand{\xmark}{\ding{55}}%

\usepackage[backend=biber, doi=true, isbn=false, url=false, eprint=false, giveninits=true, style=alphabetic, defernumbers=false, sorting=none, maxbibnames=99]{biblatex}
\title{Downscaling Neural Network for Coastal Simulations} 

\author{Zhi-Song Liu\footnote{Corresponding author: \texttt{zhisong.liu@lut.fi}} \and Markus B\"uttner \and Matthew Scarborough \and Eirik Valseth \and Vadym Aizinger \and Bernhard Kainz \and Andreas Rupp}

\date{\centering
\begin{minipage}{.9\textwidth}
\small \textbf{Zhi-Song Liu}, Lappeenranta-Lahti University of Technology LUT, Lahti, Finland
\end{minipage}\\[1ex]
\begin{minipage}{.9\textwidth}
\small \textbf{Markus Büttner, Vadym Aizinger}, University of Bayreuth, Bayreuth, Germany
\end{minipage}\\[1ex]
\begin{minipage}{.9\textwidth}
\small \textbf{Matthew Scarborough, Eirik Valseth}, Norwegian University of Life Sciences, Ås, Norway
\end{minipage}\\[1ex]
\begin{minipage}{.9\textwidth}
\small \textbf{Eirik Valseth}, Simula Research Laboratory, Oslo, Norway
\end{minipage}\\[1ex]
\begin{minipage}{.9\textwidth}
\small \textbf{Bernhard Kainz}, Friedrich-Alexander-University Erlangen-Nuremberg, Erlangen, Germany
\end{minipage}\\[1ex]
\begin{minipage}{.9\textwidth}
\small \textbf{Andreas Rupp}, Saarland University, Saarbrücken, Germany
\end{minipage}%
}

\begin{document}

\maketitle

\begin{abstract}
 Learning the fine-scale details of a coastal ocean simulation from a coarse representation is a challenging task. For real-world applications, high-resolution simulations are necessary to advance understanding of many coastal processes, specifically, to predict flooding resulting from tsunamis and storm surges. We propose a Downscaling Neural Network for Coastal Simulation (DNNCS) for spatiotemporal enhancement to learn the high-resolution numerical solution. Given images of coastal simulations produced on low-resolution computational meshes using low polynomial order discontinuous Galerkin discretizations and a coarse temporal resolution, the proposed DNNCS learns to produce high-resolution free surface elevation and velocity visualizations in both time and space. To model the dynamic changes over time and space, we propose grid-aware spatiotemporal attention to project the temporal features to the spatial domain for non-local feature matching. The coordinate information is also utilized via positional encoding. For the final reconstruction, we use the spatiotemporal bilinear operation to interpolate the missing frames and then expand the feature maps to the frequency domain for residual mapping. Besides data-driven losses, the proposed physics-informed loss guarantees gradient consistency and momentum changes, leading to a 24\% reduction in root-mean-square error compared to the model trained with only data-driven losses. To train the proposed model, we propose a coastal simulation dataset and use it for model optimization and evaluation. Our method shows superior downscaling quality and fast computation compared to the state-of-the-art methods.
\end{abstract}

\section{Introduction}
%
The two-dimensional shallow-water equations (SWE) can be used to model circulation in the global, regional, and coastal ocean, inland seas, lakes, and rivers; they are also frequently employed for atmospheric circulation studies~\cite{Kernkamp2011,DuebenKA2012,Tumolo2015}. Currently, SWE-based numerical software packages are the main tool employed in the operational forecast of tsunamis and storm surges~\cite{ADCIRC2010,Tsunawi2011,Wichitrnithed2024,ButtingerKHCSBW22}. For many coastal ocean applications---flooding simulations are one crucial example---the accuracy of the model results strongly depends on the resolution of the computational mesh and the accuracy of the time discretization. To reliably meet the requirements of accurate prediction of inundation with the purpose of warning and hazard management, mesh resolutions down to and even below 10m in the affected areas~\cite{Baptista2011} are necessary. Even using unstructured scenario-adapted meshes and GPU (Graphics Processing Unit) computing~\cite{MoralesEtAl20,Shaw2021}, such resolution requirements are computationally challenging, especially in real-time warning systems~\cite{MorauRB23}. 

Because coastal ocean simulations are important and challenging, they have been a focus of many research efforts resulting in numerous numerical and algorithmic advances in recent years. While many established software packages rely on Finite Volume or Finite Difference methods, the discontinuous Galerkin (DG) and other related discretization techniques (such as enriched Galerkin or hybridized DG methods) emerged in the last two decades as strong candidates, especially for unstructured mesh and adaptive simulations. For example, \cite{Eskilsson2004} derives an $hp$ adaptive DG discretization, \cite{Bui16,SamiiKMD19} develop hybridized DG methods and suitable time-stepping approaches, whereas \cite{HauckAFHR2020} proposes an adaptive enriched Galerkin scheme. A multiwavelet approach is explored in~\cite{Gerhard2015}, while~\cite{HajdukHAR2018} further enhances the concept of adaptivity by omitting computations in the subdomain parts, where no changes take place. 

Wetting and drying is one of the key processes in coastal regions; it represents a challenge in the numerical solution of the SWE equations because a water height of zero poses significant problems when dividing by $H$ in \eqref{momentum}. Several works~\cite{Ern2008,Bunya2009,Xing2010} propose techniques to address this challenge and the related issues of negative $H$ in DG simulations, which may result in nonphysical behavior. Accurately simulating flooding scenarios for tsunamis and storm surges relies on high-resolution meshes and very small time steps (with correspondingly high demands on computational resources~\cite{contreras2023channel,wichitrnithed2025}) further motivating techniques allowing to reduce the computational overhead.

\begin{figure}[t]\centering
 \includegraphics[width=\textwidth]{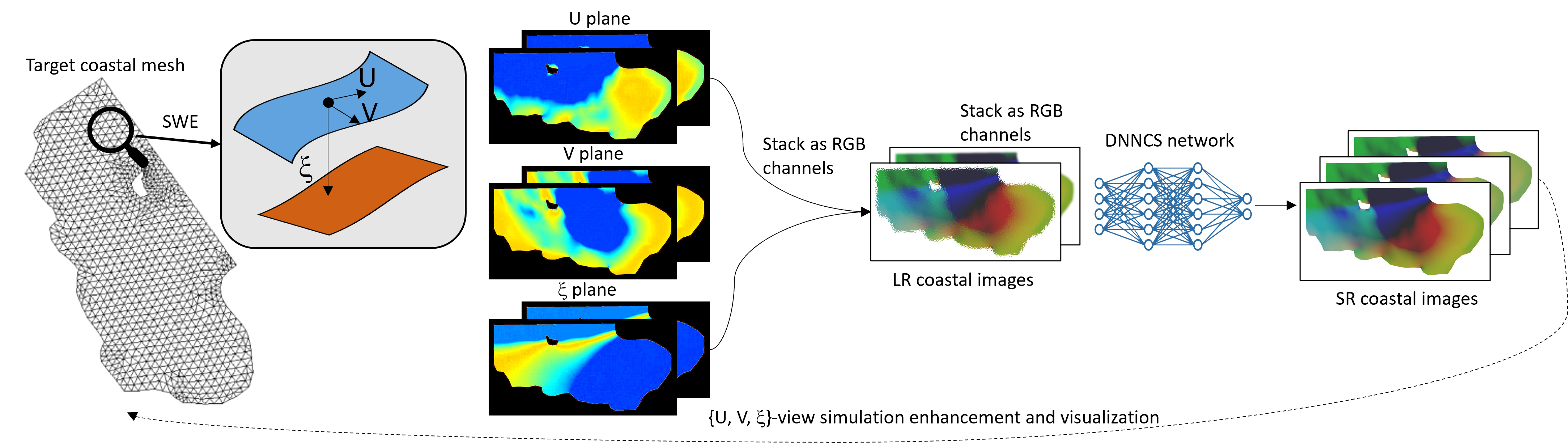}
 \caption{\small \textbf{The overall deep learning pipeline for coastal ocean downscaling.} We use the numerical model to simulate the coarse solution representation as $U, V, \xi$. We stack them as RGB channels to form a low-resolution (LR) image, which will be taken as input to the proposed \net for spatiotemporal upsampling. The obtained downscaling coastal images seamlessly enhance the spatial and temporal details, which will be useful for coastal ocean simulation and visualization.}\label{fig:teaser}
\end{figure}

Recently, there have been rapid developments of deep learning based super-resolution in image and video processing~\cite{rdn,stablesr,edvr,sr_1}. The ill-posed problems in image processing are considered to be similar to the downscaling problem in scientific modeling, like fluid dynamic simulation~\cite{shu}, climate downscaling~\cite{Teufel}, and smoke simulation~\cite{smoke}. However, there are no or very few works on applying deep learning approaches for coastal ocean downscaling. Filling this knowledge gap is the main purpose of the current study; specifically, we propose a Downscaling Neural Network for Coastal Simulations (DNNCS) to downscale the results of low-resolution (LR) coastal simulations to approximate high-resolution (HR) results of the numerical simulation. The overall process is shown in Figure~\ref{fig:teaser}. We convert the coarse simulations of the target coastal model into three views: the horizontal depth-integrated horizontal velocity field $(U,V)$ and the water height ($\xi$). Then, we merge them as RGB (Red-Green-Blue) image inputs to the proposed \net for downscaling. The super-resolved outputs can be used in place of fine-resolution simulations, e.g., to visualize the results. Besides the fast computation for HR shallow water simulations, another major advantage of our approach compared to other ML-based techniques is the fact that the coarse-resolution solution approximately satisfies physical constraints (in particular, the coarse-resolution solution is mass conservative), preventing the reconstructed fine-resolution solution from `drifting' away from physical consistency for arbitrarily long simulation runs.

To summarize, our main contributions are as follows:
\begin{itemize}[noitemsep]
\item We present a novel \net approach for coastal ocean simulation, capable of efficiently generating high-resolution data and reducing the computational burden associated with PDE-based physical modeling.
\item We introduce a spatio-temporal attention mechanism to jointly learn spatial and temporal downscaling, incorporating coordinate information as positional signals for grid-aware reconstruction.
\item To ensure physical consistency, we incorporate physics-informed loss functions as constraints, guiding the model to produce reliable high-resolution outputs.
\item We construct a new coastal simulation dataset featuring multiple levels of resolution to support effective model training and validation.
\end{itemize}

\section{Governing equations and simulation software}
%

Our governing equations are the 2D SWE in a conservative form on a 2D domain $\Omega$ given by
\begin{subequations}\label{EQ:swe}
\begin{gather}
 \partial_t \xi+\nabla \cdot \boldsymbol{q} = 0,\label{mass} \\
 \partial_t \boldsymbol{q} +\nabla \cdot \left(\tfrac{\boldsymbol{qq}^T}{H} \right) +\tau_{\textrm{bf}} \boldsymbol{q} +\left( \begin{smallmatrix} 0 & -f_c\\ f_c & 0 \end{smallmatrix} \right) \boldsymbol{q} +gH\nabla\xi =\boldsymbol{0}.\label{momentum}
\end{gather}
\end{subequations}
Here, $\boldsymbol{q} \coloneqq (U, V)^T$ denotes the depth-integrated horizontal velocity, $\xi$ the water elevation above some datum (e.g., the mean sea level), $h_b$ the bathymetric depth respective the same datum, and $H \coloneqq  h_b + \xi$ the total water depth. The remaining terms are defined as follows: $f_c$ is the Coriolis coefficient, $g$ is the gravitational acceleration, and $\tau_{\textrm{bf}}$ is the bottom friction coefficient. 

The boundary conditions relevant to our scenario are the land boundary condition $\boldsymbol{q} \cdot \boldsymbol{n} = 0$ and the open sea boundary condition $\xi = \hat{\xi}$, which prescribes the tidal elevation at open sea boundaries given as a~space- and time-dependent function $\hat \xi$.

The discretization of the SWE system defined in~\eqref{EQ:swe} using the discontinuous Galerkin (DG) method was initially realized in UTBEST~\cite{AizingerDawson2002}, a non-publicly available code developed at UT Austin which uses unstructured triangular meshes with DG polynomial approximations of orders zero (piecewise constant), one (piecewise linear), or two (piecewise quadratic). The time discretization is performed via explicit strong stability preserving Runge--Kutta methods~\cite{CockburnShuRKDG21989} of orders one (explicit Euler), two, and three chosen in accordance to the spatial discretization.

The original UTBEST scheme proved very successful and has been further developed and transferred to several code development frameworks such as Matlab/GNU Octave~\cite{HauckAFHR2020}, EXASTENCILS code generation framework~\cite{FaghihNainiKAZGK2020,AltKFFOPAHK2023,FaghihNainiKZKKA2023,FaghihNainiAKAK2025}, OpenCL~\cite{KenterSFA2021,FajKFPA2023}, or SYCL~\cite{BuettnerAKKPA2024,BuettnerAKKPA2025}. UTBEST also served as a basis for a~fully-featured 3D regional ocean model UTBEST3D~\cite{DawsonAizinger2005,AizingerPDPN2013}. 

\section{Related Works}
High-resolution coastal ocean simulation is important for accurately representing complex circulation---particularly relevant in the presence of irregular coastlines and strongly varying topography. However, simulating shallow-water equations at high spatial and temporal resolution requires massive computational efforts. To upsample the low-resolution scientific data to high-resolution full-field dynamics is called super-resolution (SR) in image and video processing, and downscaling in climate and weather prediction. Let us revisit related topics in two categories: deep learning for image/video super-resolution and physics-informed downscaling.

\subsection{Learning based super-resolution and downscaling}
Image and video super-resolution is an ill-posed problem. One LR image can lead to multiple plausible sharp and clean HR images. The typical solution is to use paired LR-HR data to optimize restoration models via pixel-based loss functions. Since the seminal super-resolution work of Dong et al.~\cite{srcnn}, many image super-resolution approaches~\cite{sr_2,sr_4,srgan,swinir,stablesr,han,tdan,duf} have adopted an end-to-end neural network to learn the regression model for reconstruction. Recently, attention~\cite{attention} has also been used in image super-resolution to involve more pixels for nonlocal pattern exploration. For example, SwinIR~\cite{swinir} proposes to use Swin Transformer~\cite{swin} for multi-scale nonlocal feature matching, HAN~\cite{han} proposes to combine spatial- and channel-wise attention to model the holistic interdependencies among layers and HAT~\cite{hat} activates more pixels for high-frequency detail reconstruction via window-based and channel attention. To overcome the bottleneck of information loss caused by the deep attention layers, \cite{drct} proposes dense-residual connections to mitigate the spatial loss and stabilize the information flow. Generative adversarial networks~\cite{gan} point to a new direction of photorealistic image super-resolution, where the model learns the manifold of natural images to produce images with pleasant visual quality. SRGAN~\cite{srgan} and ESRGAN~\cite{esrgan} are two efficient approaches that can generate super-resolved images with fine details. Following this direction, a lot of works have been exploring denoising diffusion probabilistic model (DDPMs)~\cite{ddpm}, score-based models~\cite{scorebased}, and their recent variations~\cite{consistency,cold,ide,autoregressive,ldm} for generative image super-resolution. SR3~\cite{sr3} and StableSR~\cite{stablesr} are two representative approaches that achieve photorealistic image reconstruction. However, they also suffer from high computational costs and slow inference.

Unlike image super-resolution, video super-resolution requires solving spatiotemporal reconstruction such that the resultant video has high visual quality and seamless motion changes. Temporal alignment plays an important role in temporal interpolation and frame enhancement. Optical flow is one effective approach that can estimate the motions between images and perform warping. Without knowing the ground truth optical flow, TOFlow~\cite{toflow} proposes a trainable motion estimation module to predict the motion for video super-resolution. DUF~\cite{duf} and TDAD~\cite{tdan} propose implicit motion estimation via dynamic upsampling filters and deformable alignment networks. EDVR~\cite{edvr} proposes to learn attention-based pyramid deformable convolution layers for motion estimation and then fuse multiple frames for super-resolution. Another attention-based video SR is \cite{vrt}, which encodes video frames as patch tokens and learns spatiotemporal correlation and warping for better reconstruction. Similar to image super-resolution, there are also some developments in using the diffusion model for video enhancement. \cite{upscale} proposes a training-free flow-guided recurrent module to explore latent space super-resolution. Allowing text prompts to guide texture creation can balance restoration and generation. 

With these advances in deep learning-based super-resolution, its application to climate and weather prediction has gained significant attention, namely climate/weather downscaling. That is, downscaling the grid size for ocean and weather simulation so we can get fine-resolution visualization. For example, it can be used in the sea surface temperature (SST) for high-resolution air-sea interaction study~\cite{sst,sst_2}. It can also find the links between coarse sea surface height (SSH) and high-resolution SST fields~\cite{ssh,ssh_2}, which can help to understand ocean circulation and sea surface topography. Several works~\cite{graphcast,gwm,harder_1,climax} have explored downscaling spatial resolution from 50$\sim$100 km to 1$\sim$25 km for climate data, and improving temporal resolution to hourly data. This enables more accurate regional infrastructure planning, resource adequacy evaluation, and risk assessment.

\subsection{Physics-informed downscaling/super-resolution}
Thanks to the great success in image and video super-resolution, considerable works exist about the interplay of physics and machine learning. The main trend is to approximate the HR data representation based on potentially noisy and under-resolved simulations, like smoke~\cite{smoke}, climate~\cite{graphcast,pangu}, and chemistry~\cite{material2,alphaflow}. Most deep learning-based SR for scientific data can be categorized into two groups: spatial downscaling and temporal downscaling. For the former, it is similar to image super-resolution where, given the LR data representation, the network should produce fine-grained data representation. For example, Fukami et al.~\cite{Fukami} use the SRCNN~\cite{srcnn} model to upsample 2D laminar cylinder flow. MeshfreeFlowNet~\cite{MeshfreeFlowNet} is proposed to reconstruct the turbulent flow in the Rayleigh-Benard problem via a UNet structure. Given the challenge in the laminar finite-rate-chemistry flows, \cite{piesrgan} proposes to use PIESRGAN to estimate the high-resolution flow. Using subgrid turbulent flow models, the idea is to extend ESRGAN~\cite{esrgan} to the 3D space. Two physics-informed losses, gradient loss and continuity loss, contribute to the gradient and total mass changes. PINNSR uses RRDB blocks~\cite{rdn} to build a GAN-like network to simulate the fine-grid Rayleigh-Taylor instability. The physical loss is used to govern the advection-diffusion process. PhySR~\cite{physr} proposes to use a physics-informed network to learn temporal downscaling. The key idea is to use ConvLSTM~\cite{lstm} to learn temporal refinement and dynamic evolution on LR features. Stacked residual blocks are used to learn pixel reconstruction. PhySRNet~\cite{physrnet} proposes an unsupervised learning approach to approximate the high-resolution counterparts without requiring labeled data.

Similarly, Gao et al.~\cite{gao} also propose using conservation laws and boundary conditions of fluid flows so that the model can be optimized in a self-supervised manner. Teufel et al.~\cite{Teufel} propose to predict fine-grid regional climate simulations via a Feature Split and Reconstruction (FSR) network, which can learn temporal interpolation via flow warping. \cite{shu} modifies the diffusion model to simulate the computational fluid dynamics data. It is trained only on high-resolution to learn multi-scale downsampling scenarios, resulting in a robust downscaling model for 2D turbulent flow estimation. Distinct from existing methods, ours focuses on spatiotemporal downscaling and explicitly explores the multidimensional signal via a spatiotemporal attention module. We also explore physics-informed loss to govern the optimization process and show significant improvements compared to others.

It is worth noticing that applying deep neural networks to the real physical world must produce results that guarantee conservation of key physical quantities such as mass or energy over long simulation time intervals. For instance, given paired LR and HR image pairs, the sum of values at local HR pixels needs to match with corresponding LR pixels. In deep neural networks, the common solution is using physics-informed losses. For example, Beucler et al.~\cite{beucler} propose soft penalties on the loss terms to emulate the cloud process, so that the network follows the conservation of enthalpy, the conservation of mass, terrestrial radiation, and solar radiation. The high intrinsic uncertainty of the climate system makes it difficult for neural networks to predict long-term simulations. Harder et al.~\cite{harder_2} propose a correction layer attached to the neural network and train it via physical data generated by the aerosol microphysics model. It guarantees perfect mass conservation and significant speedup. To better preserve the physical quantities, \cite{harder_1} proposes multiple local average pooling constraints to supervise the training optimization, thereby eliminating physics violations, produce no negative pixels, and preserve mass up to numerical precision. In contrast to all the above methods, we propose to use a numerical base method that guarantees the conservation of mass and momentum and conducts downscaling concerning the results of this method. Thus, we guarantee that mass and momentum are conserved over time and enable reliable long-term simulations.

\section{Approach}
This section introduces our workflow, which includes the coastal ocean simulation and the proposed downscaling neural network.

\subsection{Problem setup for coastal simulation}
Given a specific coastal region that we want to model, we first use our DG-based SWE model code for hydrodynamics simulation. Each coastal region is simulated with varying polynomial approximation orders, mesh resolutions, and time steps to capture different levels of detail and accuracy. 
The simulations are then rendered as images to facilitate the neural network training, leveraging the spatial structure of the data. Specifically, we convert the simulations into three views: the horizontal depth-integrated velocity field ($U$,$V$) and the water elevation ($\xi$), stacking them as RGB channels to create a comprehensive visual representation of the ocean dynamics. For each resolution, we employ three different modes of grid interpolation to examine the impact of interpolation methods on the simulation accuracy and neural network training effectiveness. These simulations at varying spatial and temporal resolutions are used for neural network training to enhance the model's ability to generalize across different scales and time frames.

\subsection{Proposed DNNCS for coastal downscaling}
\begin{figure*}[t] \centering
 \includegraphics[width=\textwidth]{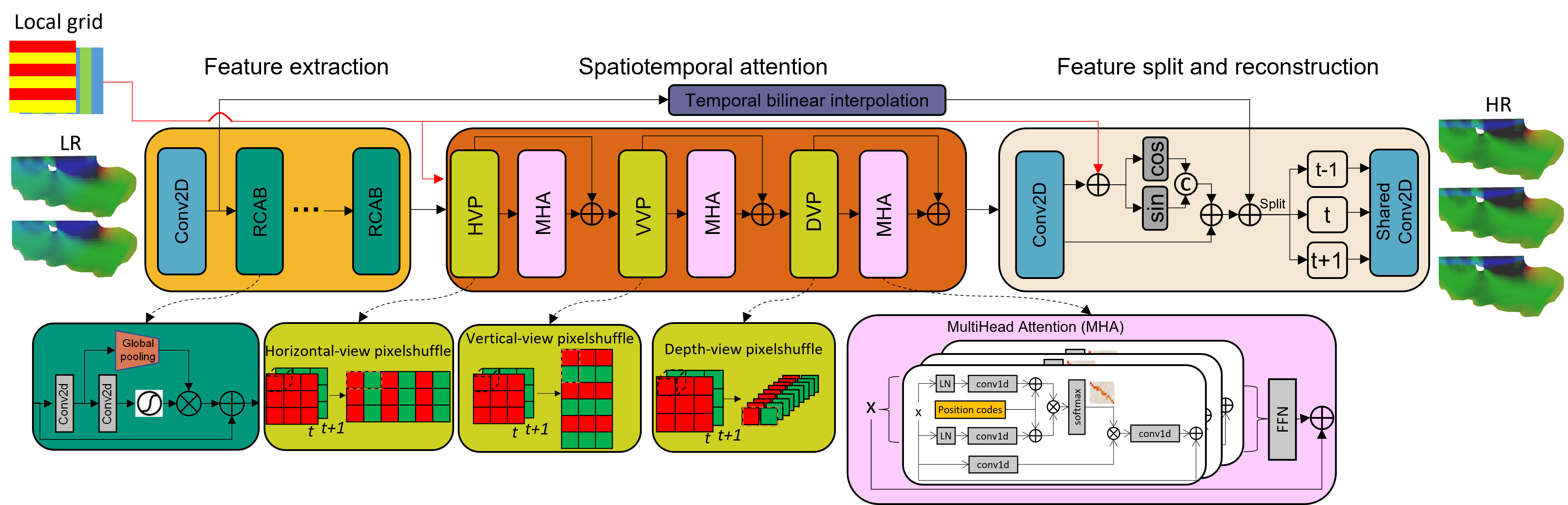}
 \caption{\small{\textbf{The overall structure of the proposed \net\!\!.} We show the complete architecture of our proposed DNNCS. Given two consecutive coastal simulations, we take them as input to first extract the deep feature representation via multiple RCAB (Residual Channel Attention Block) blocks. Then we use spatiotemporal attention to learn the pixel correlations across space and time. Finally, we split the features into three channels for temporal interpolation and spatial downscaling.}}
 \label{fig:network}
\end{figure*}

In Figure~\ref{fig:network}, we show the overall structure of our proposed \net\!\!. Given the input coarse simulation data $\mathbf{X}\in \mathbb{R}^{T\times H\times W\times C}$, where T is the number of adjacent views, H and W are the height and width, and $C=3$ is the $U$, $V$, $\xi$ planes. The proposed \net learns the mapping function $f\colon \mathbf{X}\rightarrow \mathbf{Y}$, where $\mathbf{Y}\in \mathbb{R}^{\alpha T\times H\times W\times C}$ is the high-resolution coastal estimation, and $\alpha$ is the temporal upsampling factor. Note that we do not change the spatial resolution because the low-resolution and high-resolution coastal images are rendered from the graph with the same scale but with coarse and fine mesh resolutions, respectively. That is, the resolution of the simulation video does not change, but it becomes more detailed. Inside the network are three key components: Feature extraction, Spatiotemporal attention, and Feature split and reconstruction. Let us introduce them in detail. We train DNNCS on the Bahamas dataset at different downscaling scenarios, and then fine-tune it on the Galveston dataset. The detail of the proposed Bahamas and Galveston datasets will be introduced in Section 5.1

\subsection{Feature extraction}
The backbone of the feature extraction is based on the Residual Channel Attention Network (RCAN)~\cite{rcan}, which takes the adjacent coarse inputs and jointly learns the spatial feature maps as,

\begin{equation}
 \mathbf{Z} = \sigma \left(W_2(W_1(\mathbf{X}))\right) \times GP(W_1(\mathbf{X})) + \mathbf{X}
\label{eq:rcan}
\end{equation}

\noindent where $\mathbf{Z}$ is the extracted feature, $W_1$ and $W_2$ are learnable 2D convolutional parameters, $\sigma$ is the sigmoid function and $GP$ is the global pooling operation. We stack multiple RCAN blocks to learn deep feature representations.

\subsection{Spatiotemporal attention}
\begin{figure}[t]\centering
 \includegraphics[width=0.8\columnwidth]{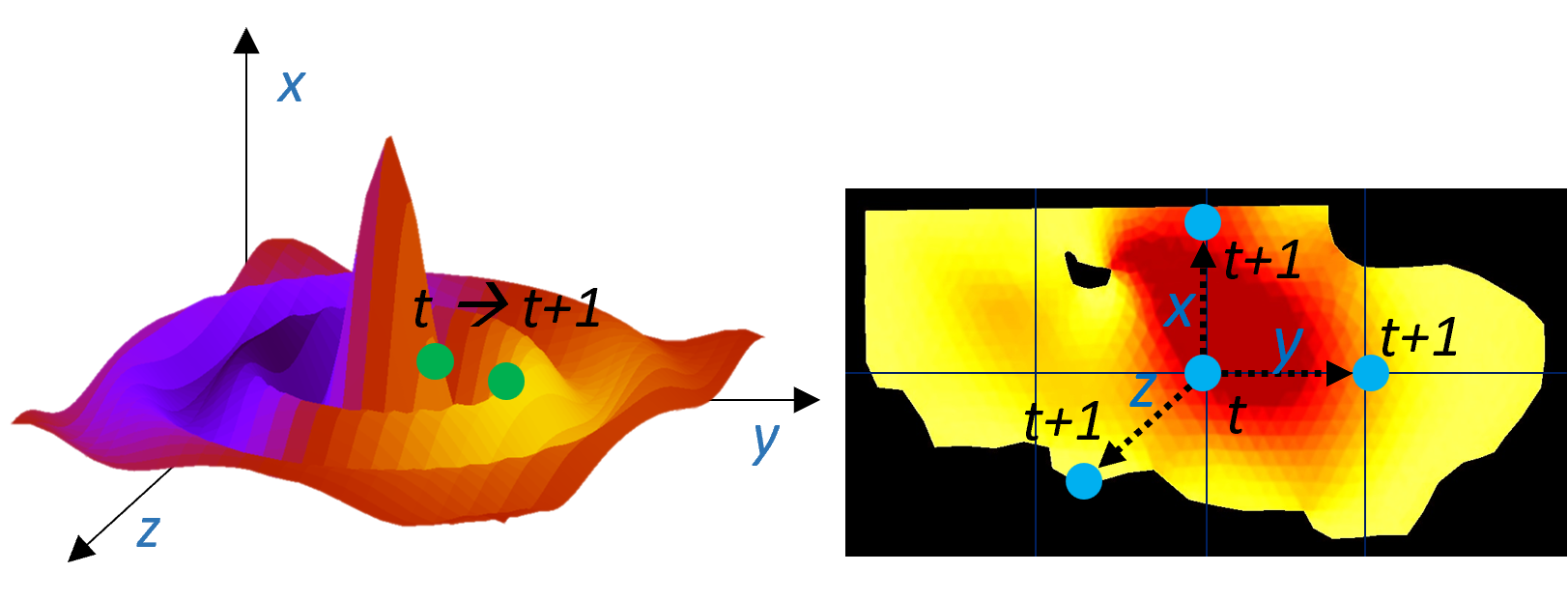}
 \caption{\small{\textbf{The spatiotemporal correlation of the water movement.} We visualize the water movement as a 3D heatmap, and we can see that the particle at the same coordinate can relate to neighborhood particles and to itself in the next step.}}
 \label{fig:dispacement}
\end{figure}
The key component of the \net is the spatiotemporal attention. Our goal is to jointly super-resolve the coarse coastal images spatially and temporally. Because PDEs govern the spatiotemporal dynamics we aim to model, we draw inspiration from the wave equation:
\begin{equation}
\frac{\partial^2 u}{\partial t^2} = c^2\left( \frac{\partial^2 u}{\partial x^2} + \frac{\partial^2 u}{\partial y^2} + \frac{\partial^2 u}{\partial z^2} \right)
\label{eq:wave}
\end{equation}
where $c$ is a fixed, positive real coefficient, and $u$ is the scalar displacement field. We can observe that the displacement acceleration is related to the spatial changes around the neighborhood. Inspired by this observation, we design the spatiotemporal attention that transfers the temporal changes across two adjacent frames to the spatial displacement. As shown in Figure~\ref{fig:dispacement}, the particle at coordinate $(i, j, k)$ can move in any direction ($x$-, $y$-, or $z$-axis) from time $t_n$ to $t_{n+1}$. Hence, we can use Pixel shuffle to conduct the sub-pixel processing in horizontal, vertical, and depth dimensions.

Mathematically, we can describe the process as
\begin{subequations}\label{eq:sta}
\begin{align}
 H(\mathbf{Z}) & = \mathbf{Z} + G\left(r_h(\mathbf{Z}) + t_h(\mathbf{p})\right), \\
 V(\mathbf{Z}) & = H(\mathbf{Z}) + G\left(r_v(H(\mathbf{Z})) + t_v(\mathbf{p})\right), \\
 D(\mathbf{Z}) & = V(\mathbf{Z}) + G\left(r_d(V(\mathbf{Z})) + t_d(\mathbf{p})\right),
\end{align} 
\end{subequations}
where $r_{\{h,v,d\}}$ are the horizontal, vertical, and depth pixel shuffle operators. It fuses the $t_n$-th and $t_{n+1}$-th wave equation features into one spatiotemporal map. $H(\cdot)$, $V(\cdot)$ and $D(\cdot)$ are the corresponding spatial maps that fuse the temporal features along the three dimensions. $G(\cdot)$ is the MultiHead Attention (MHA) operation. The learnable parameters are shared across horizontal-, vertical- and depth-view computation, which can find the nonlocal correlations across space and time. As shown in Figure~\ref{fig:dispacement}, we fuse the features at adjacent frames to calculate the correlations among different directions. In order to record the pixel positional information, we propose to use learnable positional codes \textbf{p} to record the relative feature positions and attach them to the corresponding feature maps for computation.

MultiHead Attention (MHA) is a standard nonlocal feature extraction that groups feature maps into subsets and computes the attention map in each subset in parallel. Let us denote the input feature as X; we describe the MHA process as,
\begin{equation}
\begin{matrix}
\!\begin{aligned}
    & Y = FFN(MHA(X)) + X \\
    & MHA(X) = X + Concat(head_1, head_2, ..., head_n)W_O \\
    & head_i = softmax\left( \frac{LN(W_Q(X))LN(W_K(X))}{\sqrt{d}}\right)W_V(X)
\label{eq:attn}
\end{aligned}
\end{matrix} 
\end{equation}
where $Concat$ concatenates the subset of features into one, $W_{\{Q,K,V\}}$ are query, key and value matrices. $W_O$ is the weighting matrix for total $O$ heads. $LN$ is the layer normalization process, $FFN$ is the feed-forward network with multiple convolution layers for the attention output, and $d$ is the dimension of the feature map, which is used for normalization.

\subsection{Feature split and reconstruction} 
In the feature split and reconstruction process, we first project the learned features into the frequency domain using cosine and sine operators\footnote{Note that we use cosine and sine operators for computational efficiency as they are used often in image/video coding to capture high-frequency details.}, then combine these components to enhance the original features. We subsequently apply a simple convolution layer to extend these features and split them into three subsets, each representing one of the three temporal feature maps for the coastal data. The input consists of two low-resolution coastal images, with the proposed network (\net\!\!\!) tasked with predicting the intermediate image. To achieve this, we use bilinear interpolation for temporal frame interpolation on the initial feature maps. A short connection is implemented to ensure that the spatiotemporal attention mechanism effectively learns the residuals. Finally, a shared convolution operation is employed to directly output the predicted downscaling coastal images. This approach ensures that the model captures both spatial and temporal details, producing high-quality intermediate frames. 

\subsection{Losses for coastal downscaling}
To train the proposed \net\!\!\!, we propose utilizing three loss terms to better constrain the visual consistency in space and time, including MAE (Mean Absolute Errors) loss, $L_p$ loss, and physics-informed loss. Given the ground truth and estimated downscaling results ${\mathbf{Y}, \mathbf{Y'}}\in \mathbb{R}^{\alpha T\times H\times W\times C}$, we first have the MAE loss as,
\begin{equation}
L_{mae} = \frac{1}{H\times W\times C\times T} \sum_i^H \sum_j^W \sum_k^C \sum_t^T |Y_{i,j,k}(t)-Y'_{i,j,k}(t)|
\label{eq:mae}
\end{equation}

We also need to compute the $L_p$ loss as a batch-wise weighted loss function that can balance the sample reconstruction quality. Mathematically we have, 
\begin{equation}
L_{lp} = \frac{1}{H\times W\times C\times T} \sum_i^H \sum_j^W \sum_k^C \sum_t^T \frac{\left\|Y_{i,j,k}(t)-Y'_{i,j,k}(t)\right\|^2}{\left\|Y(t)\right\|^2}
\label{eq:lp}
\end{equation}

Finally, we propose to use differential loss to compute the first-order gradient differences between ground truth and estimations. The idea is to calculate the gradient along horizontal and vertical directions. In the meantime, we also calculate the gradient along the z-axis, which represents the U-, V-, and $\xi$-planes. 
\begin{equation}
L_{diff} = \left\|\frac{dY(t)}{dx}-\frac{dY'(t)}{dx}\right\|^2 + \left\|\frac{dY(t)}{dy}-\frac{dY'(t)}{dy}\right\|^2 + \left\|\frac{dY(t)}{dz}-\frac{dY'(t)}{dz}\right\|^2
\label{eq:diff}
\end{equation}

The total loss is defined as the weighted sum of all three losses as $L=\alpha_{mae}L_{mae} + \alpha_{lp}L_{lp} + \alpha_{of}L_{diff}$. In our experiments, we set $\alpha_{mae}=4, \alpha_{lp}=1, \alpha_{diff}=100$ to balance their contributions to the network optimization.

\section{Experiments}

\subsection{Tidal scenarios}
\begin{figure}[t]\centering
 \includegraphics[width=\textwidth]{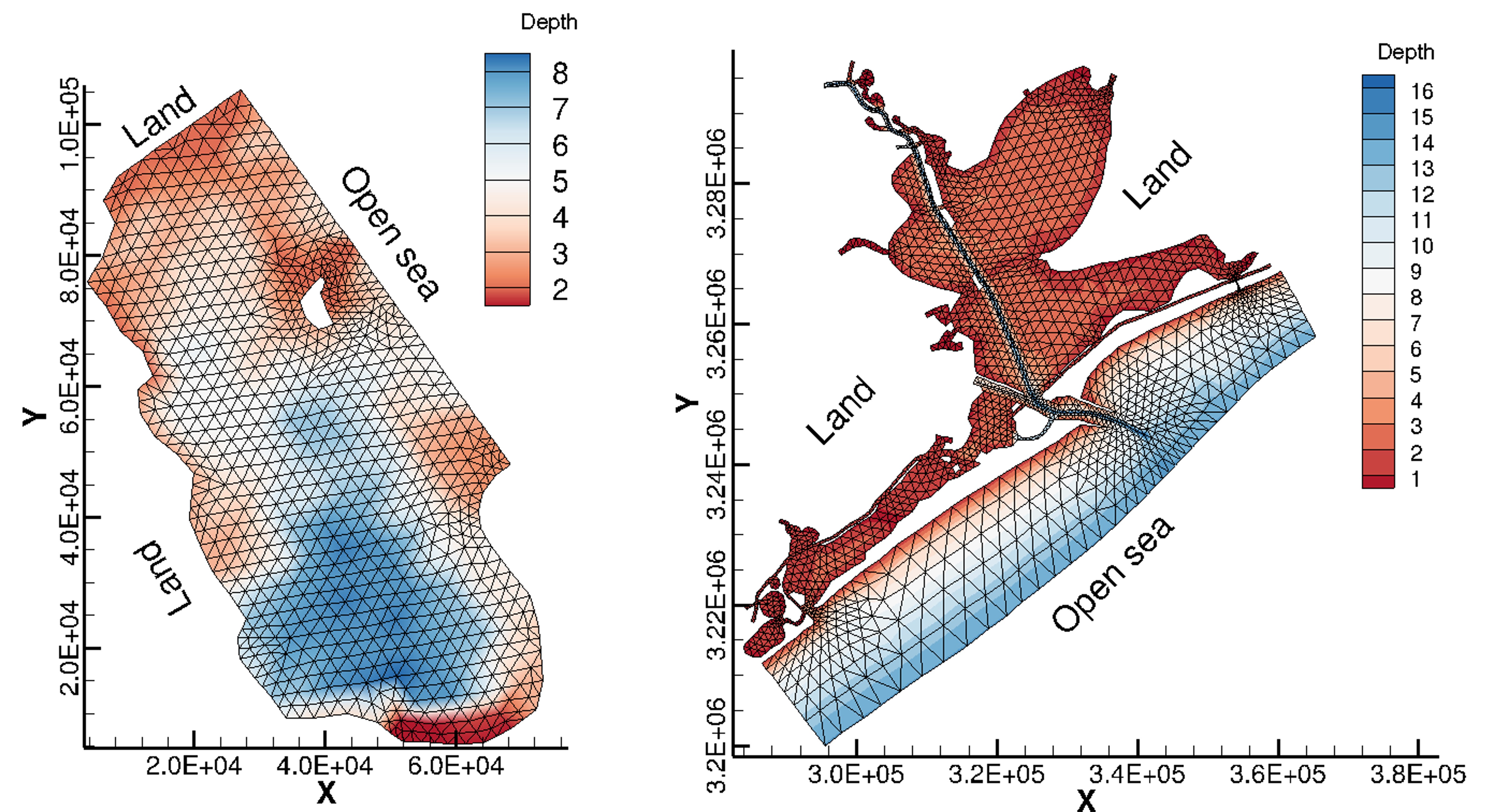}
 \caption{\small{Computational domain, bathymetry (in meters), and the coarse mesh for Bahamas (left, 1696 elements) and Galveston (right, 3397 elements) test cases. The x-axis points East and the y-axis points North.}}
 \label{fig:testcases}
\end{figure}

\paragraph{\textbf{Datasets}}
To generate sufficient data samples for model training and analysis, we use the SYCL implementation~\cite{BuettnerAKKPA2024, BuettnerAKKPA2025} of the UTBEST \cite{AizingerDawson2002} model to simulate the tidal circulation in two locations: Bahamas (Bight of Abaco) and Galveston Bay. The geometry of both computational domains, bathymetry (bottom topography), and the coarsest computational meshes are illustrated in Figure~\ref{fig:testcases}; all tidal setups in this section are based on the validated ADCIRC~\cite{adcirc} datasets. Finer meshes were obtained by subdividing each triangle into four via edge bisection, whereas the bathymetry on finer meshes was obtained by linear interpolation from the coarser mesh nodes to preserve the consistency of the problem setup between different mesh resolutions. 

The Bahamas test problem setup is based on \cite{Westerink1989}; it uses the tidal elevation consisting of five constituents (O1, K1, N2, M2, S2) at the open sea boundary, a~quadratic bottom friction $\tau_{bf} = C_f |\mathbf{q}| / H^2$ with $C_f = 0.009$, and a constant Coriolis force with the coefficient set to $3.19 \times 10^{-5}~\mathrm{s}^{-1}$. The Galveston test also uses a~quadratic friction with $C_f = 0.004$ and a constant Coriolis force with the coefficient $7.07 \times 10^{-5}~\mathrm{s}^{-1}$. In the following sections, the open boundary uses exactly the same forcing as the Bahamas test case, while in Section~\ref{sec:Galveston_new_forcing} we generate tidal components M2, N2, O1, S2, K1, P1, Q1, K2 to see the effect of changing forcing on the prediction. See~\cite{AizingerDawson2002} for a more detailed description of the test problems.

The simulations are run for 24 hours with the time step for the lowest approximation order (piecewise constant DG spaces) and the lowest mesh resolution set to 20s for the Bahamas and 12s for the Galveston dataset. Output files with the water height $\xi$ and depth-integrated horizontal velocities $U$ and $V$ are generated in each time step for all configurations. Increasing the mesh resolution or the polynomial discretization order requires smaller time steps (we halve the time step for each mesh refinement and each order increment) and results in more output files. The discontinuous Galerkin scheme uses Local Lax-Friedrichs flux and explicit strong stability preserving (SSP) Runge-Kutta time stepping schemes~\cite{Gottlieb2001} of orders one, two, and three, chosen corresponding to the order of the polynomial discretization in space.

Table~\ref{tab:data} summarizes the data used for training, with the number of mesh elements shown in ``Resolution'' and the number of time steps in the ``Order'' column. Given three approximation orders and three different resolutions, we have a total of $3\times3$ simulation cases for each location. Then, we render all data of their three views ($U$, $V$, $\xi$ planes) into grayscale images. Finally, all three views are merged as RGB images (this is purely a trick for efficient data representation and to leverage standard multi-channel convolutional architectures) as the training data. We train one model on both the Bahamas and Galveston datasets for our evaluation.

We treat the simulations from each region–order combination as an independent dataset. The proposed model is jointly optimized using all these datasets, allowing it to learn a common 4$\times$ spatial and 2$\times$ temporal downscaling across regions and orders. For example, in the Bahamas dataset, training pairs include 1696$\rightarrow$6784, and 6784$\rightarrow$27136, while in the Galveston dataset, pairs include 3397$\rightarrow$13588, and 13588$\rightarrow$54352. These spatial transitions are combined with three polynomial orders (0, 1, and 2). A single model is trained jointly on data from both the Bahamas and Galveston datasets at three polynomial orders, enabling it to learn a resolution-agnostic spatiotemporal downscaling behavior that generalizes across mesh sizes, discretization orders, and coastal regions.

\begin{table}[t]
\caption{\small\textbf{Summary of the data used in coastal simulation for Bahamas and Galveston.} The ``Elem.'' value represents the number of mesh elements, 'Min/Max' column contains the minimum/maximum element size in meters. In the ``Order'' column, we list the number of time steps in the 24-hour simulation for the corresponding polynomial discretization order, which refers to the polynomial degree used to approximate each solution component per mesh cell.}
\resizebox{\columnwidth}{!}{
\begin{tabular}{ccccc|ccccc}
\toprule 
\multicolumn{5}{c|}{Bahamas}                               & \multicolumn{5}{c}{Galveston}                              \\ \midrule
\multicolumn{1}{c|}{Elem.} & Min/Max & Order 0 & Order 1 & Order 2 & \multicolumn{1}{c|}{Elem.} & Min/Max & Order 0 & Order 1 & Order 2 \\ \midrule
\multicolumn{1}{c|}{1696}  & 640/2400  & 4320   & 8640   & 17280  & \multicolumn{1}{c|}{3397}       & 200/4400  & 1800   & 3600   & 7200   \\
\multicolumn{1}{c|}{6784}  &  320/1200   & 8640   & 17280  & 34560  & \multicolumn{1}{c|}{13588}      & 100/2200  & 3600   & 7200   & 14400  \\
\multicolumn{1}{c|}{27136}  & 160/600  & 17280  & 34560  & 69120  & \multicolumn{1}{c|}{54352}      & 50/1100  & 7200   & 14400  & 28800  \\ \bottomrule
\end{tabular}
}
\label{tab:data}
\end{table}

We randomly split the data into training, validation, and testing in a $6:2:2$ ratio without overlapping. The rendered images are of identical dimensions, approximately $900\times 500$ pixels. We also mask out the land area and only focus on the water-covered subdomain downscaling. In the training phase, we learn 4$\times$ spatial downscaling and 2$\times$ temporal downscaling across different orders. We randomly select two adjacent LR coastal images ($\mathbf{X}(t_n), \mathbf{X}(t_{n+1})$) from lower resolutions to predict three corresponding HR images $\mathbf{Y}(t_{2n}), \mathbf{Y}(t_{2n+1}), \mathbf{Y}(t_{2n+2})$ from next higher-resolution. We then divide each pair of images into smaller $64\times 64$ patches. To increase the diversity of training examples, we randomly flip them horizontally or vertically, rotate them, and reverse the order of frames, so the model sees a wider variety of patterns during training.

\paragraph{\textbf{Parameter setting}}
We train \net using Adam optimizer with the learning rate of $1\times10^{-4}$. The learning rate is halved after 30k iterations. The batch size is set to 24, and \net is trained for 100k iterations (about 16 hours) on a PC with one NVIDIA V100 GPU using the PyTorch deep learning platform. 

\paragraph{\textbf{Metrics and evaluation}}
We evaluate different models separately for two scenarios: Bahamas and Galveston. In each scenario, we average the model predictions at different orders. We use five metrics for evaluation: 
\begin{itemize}[noitemsep]
\item \textbf{MSE} (Mean Squared Error) and \textbf{MAE} (Mean Absolute Error) measure the average pixel differences between ground truth and estimation
\item \textbf{SSIM}~\cite{ssim} (Structural SIMilarity) measures the structural similarity between ground truth and estimation 
\item \textbf{GMSD}~\cite{gmsd} (gradient Magnitude Similarity Deviation) measures the gradient differences for perceptual quality assessment 
\item \textbf{LPIPS}~\cite{LPIPS} (Learned Perceptual Image Patch Similarity) measures the perceptual similarity between ground truth and estimation
\end{itemize}
To measure the model complexity, we use FLOPs (floating point operations), memory, and the number of parameters to compare the computation efficiency.

\subsection{Comparison of the spatiotemporal downscaling with state-of-the-art methods}\label{sota}
To show the efficiency of our proposed method, we compare it with four state-of-the-art image super-resolution methods: VDSR~\cite{vdsr}, RDN~\cite{rdn}, StableSR~\cite{stablesr} and SwinIR~\cite{swinir}, two video super-resolution methods: EDVR~\cite{edvr} and VRT~\cite{vrt}, and one physics-informed downscaling method: PhySR~\cite{physr}. Note that image super-resolution cannot conduct frame interpolation, so we apply spatiotemporal Bicubic to interpolate the missing frames and then apply frame-by-frame downscaling. All approaches are reimplemented using the publicly available codes, and the methods marked with $*$ indicate that they were retrained from our coastal dataset for a fair comparison. In Table~\ref{tab:sota}, both Galveston and Bahamas have two downscaling schemes. The ``Resolution'' column lists the number of grid elements. The result for each scheme is the average of all three polynomial approximation orders. We can see that our approach achieves the best performance with respect to the RMSE, MAE, SSIM, and GMSD metrics and is the second-best in LPIPS.

\begin{table}[h]\centering
\caption{\small \textbf{Comparison with state-of-the-art methods on coastal downscaling.} We report the deviations from the ground truth given by the highest accuracy simulations for the test datasets of Bahamas and Galveston using different metrics.} 
\renewcommand\arraystretch{1.6}
\resizebox{\textwidth}{!}{
\begin{tabular}{c|c|cccccccccc}
\toprule
\multicolumn{1}{r|}{Dataset} & \multicolumn{1}{r|}{Resolution} & \multicolumn{1}{r}{Method} & \multicolumn{1}{r}{ST-Bicubic} & \multicolumn{1}{r}{VDSR*} & \multicolumn{1}{r}{RDN*} & \multicolumn{1}{r}{ResShift} & \multicolumn{1}{r}{SwinIR} & \multicolumn{1}{r}{EDVR*} & \multicolumn{1}{r}{VRT} & \multicolumn{1}{r}{PhySR*} & \multicolumn{1}{r}{\cellcolor{mistyrose}{Ours}} \\ \midrule
\multirow{10}{*}{Bahamas} & \multirow{5}{*}{1696$\times$6784} & RMSE$\downarrow$ & 0.077 & 0.079 & 0.054 & 0.066 & 0.045 & 0.036 & 0.067 & 0.084 & \cellcolor{mistyrose}{0.028} \\
 &  & MAE$\downarrow$ & 0.035 & 0.033 & 0.023 & 0.041 & 0.026 & 0.014 & 0.038 & 0.036 & \cellcolor{mistyrose}{0.011} \\
 &  & SSIM$\uparrow$ & 0.964 & 0.960 & 0.971 & 0.977 & 0.975 & 0.979 & 0.978 & 0.966 & \cellcolor{mistyrose}{0.981} \\
 &  & GMSD$\downarrow$ & 0.0413 & 0.045 & 0.046 & 0.046 & 0.045 & 0.038 & 0.047 & 0.040 & \cellcolor{mistyrose}{0.033} \\
 &  & LPIPS$\downarrow$ & 0.108 & 0.118 & 0.125 & 0.133 & 0.122 & 0.119 & 0.135 & 0.113 & \cellcolor{mistyrose}{0.100} \\
 & \multirow{5}{*}{6784$\times$27136} & RMSE$\downarrow$ & 0.040 & 0.004 & 0.021 & 0.023 & 0.020 & 0.015 & 0.024 & 0.051 & \cellcolor{mistyrose}{0.014} \\
 &  & MAE$\downarrow$ & 0.0208 & 0.018 & 0.009 & 0.010 & 0.007 & 0.007 & 0.011 & 0.022 & \cellcolor{mistyrose}{0.006} \\
 &  & SSIM$\uparrow$ & 0.979 & 0.975 & 0.987 & 0.988 & 0.987 & 0.989 & 0.988 & 0.979 & \cellcolor{mistyrose}{0.990} \\
 &  & GMSD$\downarrow$ & 0.019 & 0.039 & 0.026 & 0.025 & 0.024 & 0.020 & 0.027 & 0.037 & \cellcolor{mistyrose}{0.011} \\
 &  & LPIPS$\downarrow$ & 0.097 & 0.105 & 0.105 & 0.112 & 0.107 & 0.106 & 0.114 & 0.099 & \cellcolor{mistyrose}{0.093} \\ \hline
\multirow{10}{*}{Galveston} & \multirow{5}{*}{3397$\times$13588} & RMSE$\downarrow$ & 0.039 & 0.020 & 0.021 & 0.023 & 0.024 & 0.009 & 0.023 & 0.032 & \cellcolor{mistyrose}{0.007} \\
 &  & MAE$\downarrow$ & 0.002 & 0.012 & 0.007 & 0.012 & 0.014 & 0.010 & 0.011 & 0.011 & \cellcolor{mistyrose}{0.002} \\
 &  & SSIM$\uparrow$ & 0.972 & 0.990 & 0.989 & 0.989 & 0.989 & 0.988 & 0.990 & 0.988 & \cellcolor{mistyrose}{0.996} \\
 &  & GMSD$\downarrow$ & 0.043 & 0.023 & 0.036 & 0.037 & 0.035 & 0.033 & 0.035 & 0.029 & \cellcolor{mistyrose}{0.014} \\
 &  & LPIPS$\downarrow$ & 0.108 & 0.046 & 0.033 & 0.040 & 0.027 & 0.027 & 0.037 & 0.026 & \cellcolor{mistyrose}{0.016} \\
 & \multirow{5}{*}{13588$\times$54352} & RMSE$\downarrow$ & 0.062 & 0.050 & 0.030 & 0.032 & 0.030 & 0.028 & 0.034 & 0.041 & \cellcolor{mistyrose}{0.018} \\
 &  & MAE$\downarrow$ & 0.016 & 0.016 & 0.010 & 0.014 & 0.010 & 0.009 & 0.012 & 0.014 & \cellcolor{mistyrose}{0.005} \\
 &  & SSIM$\uparrow$ & 0.979 & 0.981 & 0.984 & 0.986 & 0.986 & 0.986 & 0.987 & 0.984 & \cellcolor{mistyrose}{0.991} \\
 &  & GMSD$\downarrow$ & 0.079 & 0.059 & 0.049 & 0.047 & 0.047 & 0.049 & 0.041 & 0.045 & \cellcolor{mistyrose}{0.035} \\
 &  & LPIPS$\downarrow$ & 0.079 & 0.048 & 0.043 & 0.048 & 0.045 & 0.043 & 0.040 & 0.038 & \cellcolor{mistyrose}{0.028} \\ \bottomrule
\end{tabular}%
}
\label{tab:sota}
\end{table}

\begin{figure}[!th]
 \centering
\includegraphics[width=\textwidth]{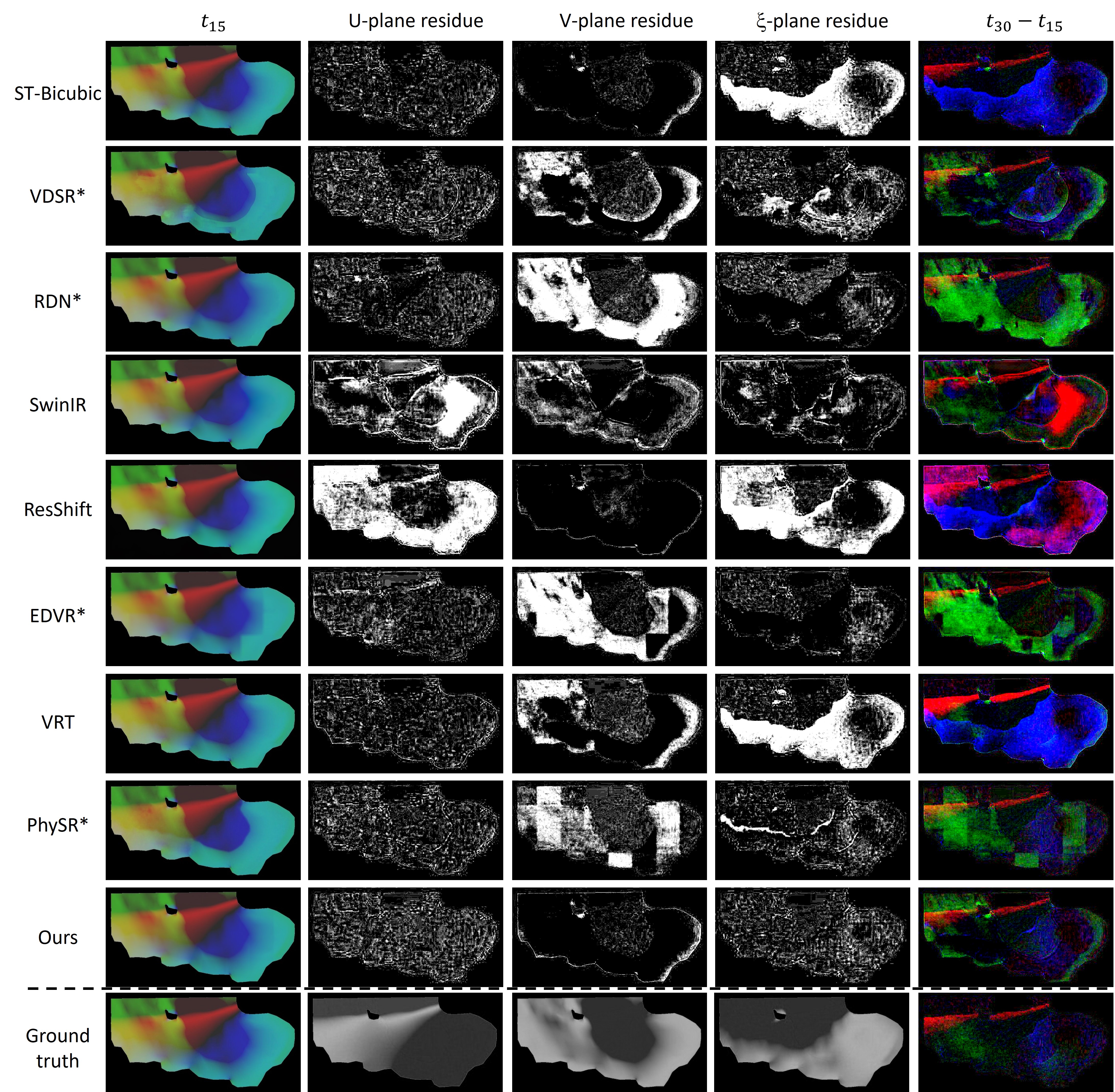}
 \caption{\small \textbf{Visual comparison among different methods on Bahamas dataset.} We show the ground truth from resolution 1696, order 0 at time $t_{15}$ and $t_{30}$, and the corresponding residual maps between prediction and ground truth (U, V, and $\xi$ planes -- we multiply the residuals by factor 50 to highlight the differences), and the residual map between the ground truths at times $t_{30}$ and $t_{15}$ (we multiply the residuals by factor 20 for visualization).}
 \label{fig:bahamas_sota}
\end{figure}

\begin{figure}[!th]\centering
 \includegraphics[width=\textwidth]{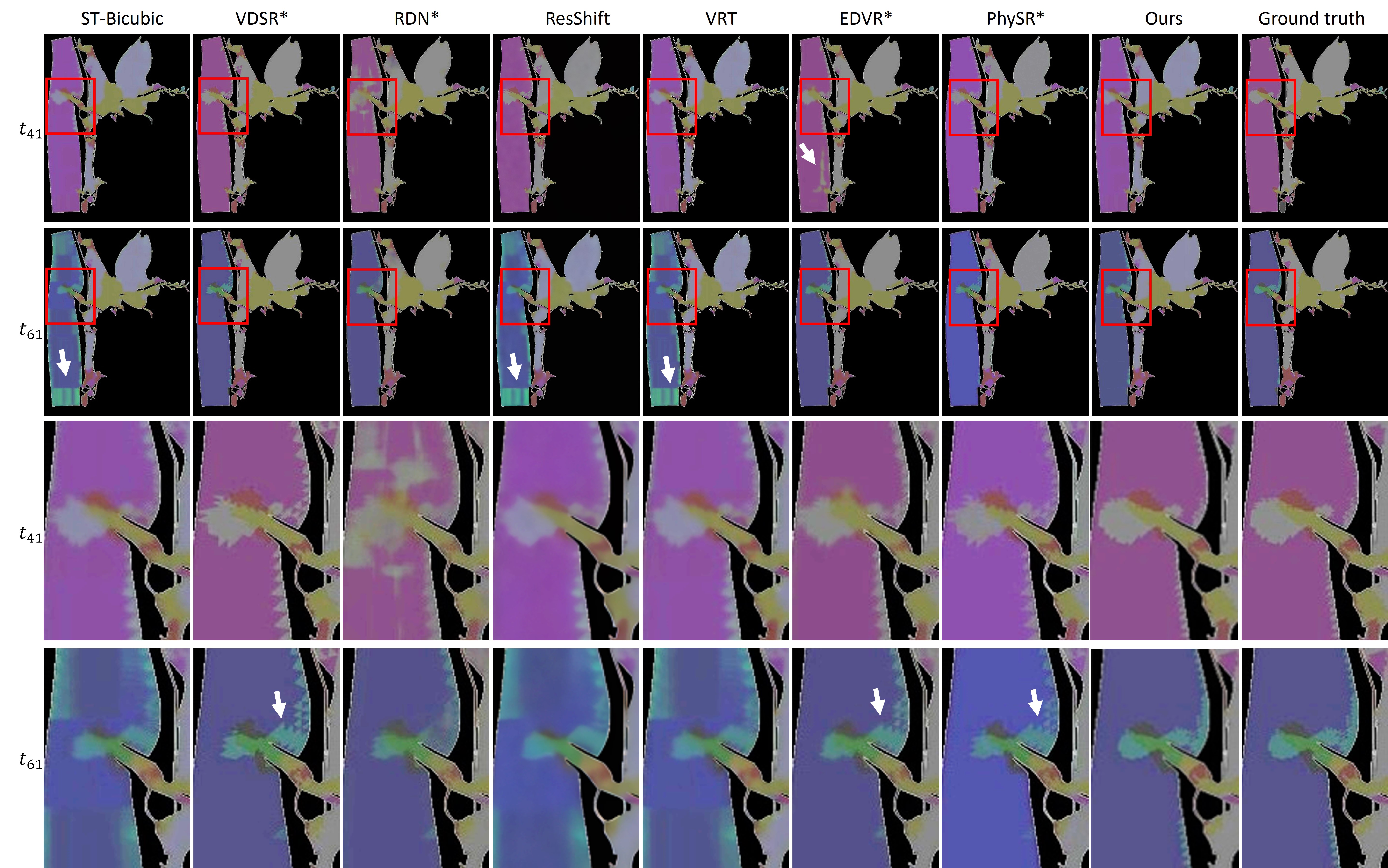}
 \caption{\small \textbf{Visual comparison among different methods on the Galveston dataset.} We use the data sample from resolution 3397, order 1 at time $t_{41}$ and $t_{61}$. The simulation results at time $t_{41}$ and $t_{61}$ in RGB images. We enlarge the red box region in rows 3 and 4 to highlight the visual differences.}
 \label{fig:galveston_sota}
\end{figure}

Visually, we report the downscaling results of different methods in Figure~\ref{fig:bahamas_sota} for the Bahamas dataset and in Figure~\ref{fig:galveston_sota} for the Galveston Bay. Each plane is normalized between [0, 255] and shown as a single-channel grayscale image, and we also stack normalized three planes $U, V, \xi$ in order as RGB images.

That is, the velocity components \( U \), \( V \), and surface elevation \( \xi \) are independently normalized for neural network training. The RGB and grayscale visualizations are therefore normalized representations used to compare prediction accuracy, rather than direct physical quantities. In the RGB visualizations, each color channel corresponds to one variable (\( U \), \( V \), or  \( \xi \)). A red-, green-, or blue-dominated pixel indicates a relatively larger error (or value) in the corresponding variable after normalization. Similarly, in the grayscale plots, brighter pixels indicate larger normalized magnitudes or errors. These colors are intended to highlight spatial error patterns and relative performance, not to convey absolute physical values.

In Figure~\ref{fig:bahamas_sota}, the last row shows the ground truth. Denoting the simulation output for the n-th time step by $t_n$, we show the results for $t_{15}$ in the first column and the residual maps of $U$-, $V$- and $\xi$-planes between predictions and ground truth in columns 2, 3 and 4, respectively. The last column illustrates the residual maps between $t_{15}$ and $t_{30}$ to visualize the temporal changes. We can observe that 1) VDSR, RDN, EDVR, and PhySR wrongly predict the $V$-plane simulations, ST-Bicubic, SwinIR, and VRT mispredict the $U$-plane simulations, ResShift fails on both U- and $\xi$-planes. We can better minimize the simulation differences simultaneously for $U$-, $V$- and $\xi$-planes. 2) From the last column, we can see that our method can better match with the ground truth changes, while other methods produce large errors. In Figure~\ref{fig:galveston_sota} of the downscaling results on Galveston Bay (here, the ground truth is shown in the last column), we use white arrows to highlight the significant differences. We can see that 1) globally, ResShift and VRT enlarge the errors caused by ST-Bicubic (the green areas around the boundaries at time $t_{61}$). EDVR produces the wrong pattern at time $t_{41}$. 2) From the enlarged areas, we can see that our method can better mimic the changes of the solution around the corners. Others, like VDSR and PhySR* magnify the zigzag patterns. ResShift and VRT oversmooth the solution which visually causes color shifts. 

Three main reasons that other methods fail while ours work: 1) Other image-based approaches (SwinIR, ResShift) are trained with additional perceptual reconstruction losses, which can cause false features or hallucinate detail. 2) Video-based approaches (EDVR, VRT) use estimated optical flow generated by pre-trained neural networks for temporal interpolation. The errors caused by optical flow can be accumulated for spatial downscaling. 3) All other works do not consider physics-informed losses to ensure physical consistency, which would violate the energy flow between frames.

\begin{figure}[h]\centering
 \includegraphics[width=0.8\textwidth]{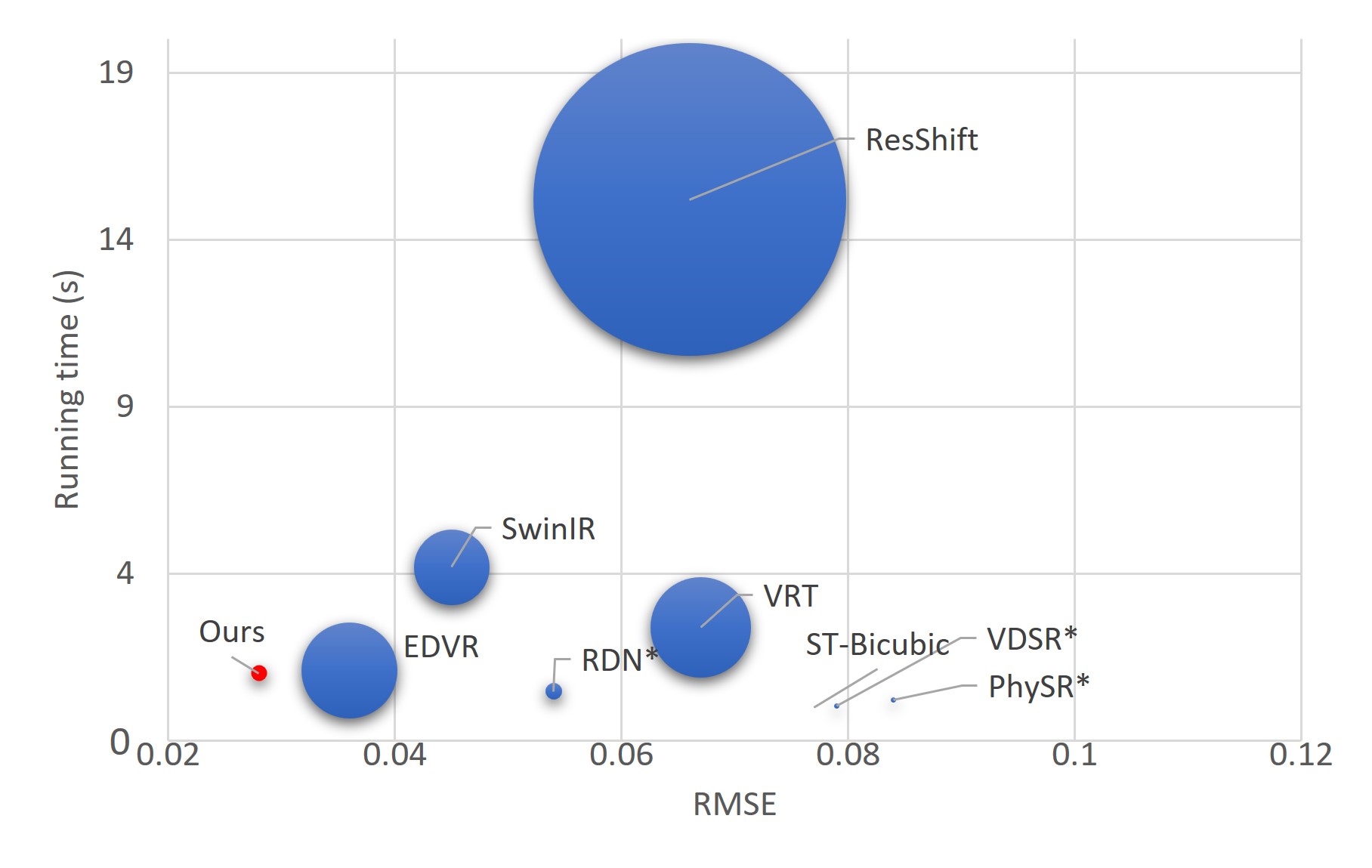}
 \caption{\small{\textbf{Speed and performance comparison.} Overall comparison among different methods in terms of speed (runtime on the vertical axis), performance (RMSE on the horizontal axis), and model complexity (size of the bubbles).}}
 \label{fig:tradeoff}
\end{figure}

We report the number of parameters, FLOPs, and runtime for model complexity in Table~\ref{tab:complexity}. The FLOPs are computed with the input of LR size $180 \times 180$ and $\times4$ upsampling settings for all comparisons. We can see that ours has the third-lowest model complexity regarding the number of parameters and runtime. In general, the model performance and complexity contradict each other. To better visualize the trade-off comparisons among different methods, we summarize the information in Table~\ref{tab:sota} and Table~\ref{tab:complexity} and visualize them in Figure~\ref{fig:tradeoff}. Ours is located at the bottom left corner, which indicates that it achieves the best balance between model complexity and reconstruction quality.

\begin{table}[!th]\centering
\caption{\small \textbf{Comparison with state-of-the-art methods on model complexity.} We report the results on the Bahamas dataset under the same hardware settings.} 
\renewcommand\arraystretch{1.4}
\resizebox{\columnwidth}{!}{
\begin{tabular}{c|ccccccccc}
\toprule
Method                   & ST-Bicubic & VDSR* & RDN*  & ResShift & SwinIR  & EDVR*  & VRT     & PhySR* & \cellcolor{mistyrose}{Ours}  \\ \midrule
Number of parameters (M) & -          & 0.651 & 21.99 & 118.59   & 11.72   & 20.6   & 35.5    & 0.817  & \cellcolor{mistyrose}{8.07}  \\
FLOPs (GMac)             & -          & 8.23  & 90.14 & 32156    & 1878.41 & 3020.1 & 3305.75 & 9.20   & \cellcolor{mistyrose}{65.15} \\
Runtime (s)              & 0.003      & 0.038 & 0.48  & 15.2     & 4.2     & 1.10   & 2.4     & 0.23   & \cellcolor{mistyrose}{1.02}  \\ \bottomrule
\end{tabular}%
}
\label{tab:complexity}
\end{table}

\subsection{Ablation studies and analysis}
\begin{figure}[!th]\centering
 \includegraphics[width=\textwidth]{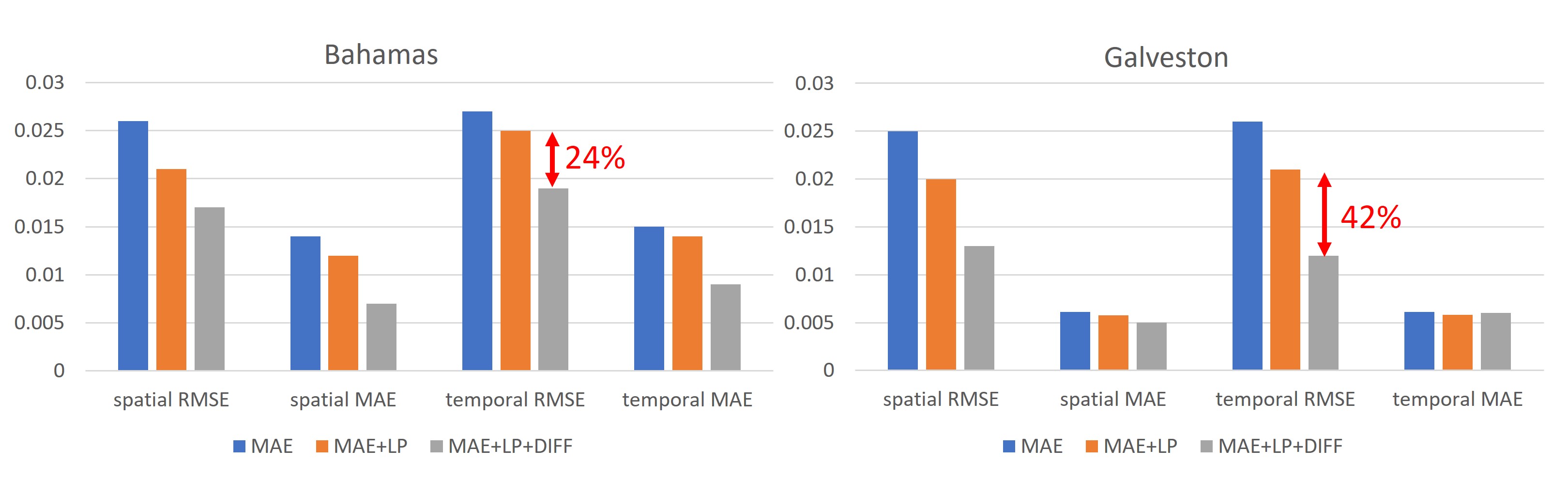}
 \caption{\small \textbf{Quantitative comparisons of using different loss terms.} We show the coastal simulation results on the Bahamas and Galveston by computing the spatial differences (pixel loss on intra frames) and temporal differences (pixel loss on inter frames).}
 \label{fig:loss}
\end{figure}

\paragraph{\textbf{Losses for optimization}} 
We propose to use three loss terms to optimize the proposed model. To show their effects, we visualize their effect by plotting the quantitative metrics in Figure~\ref{fig:loss}. Given LR input at $t_n$ and $t_{n+1}$ and the predictions at $t_{2n}$, $t_{2n+1}$, and $t_{2n+2}$, we take the coastal simulations at $t_{2n}$ and $t_{2n+2}$ (we refer to them as intra frames) to compute the spatial differences, and the coastal simulation at $t_{2n+1}$ (we refer to them as inter frames) to compute the temporal differences. We can see that using $L_{lp}$ and $L_{diff}$ losses can reduce both RMSE and MAE losses. Specifically, using $L_{diff}$ can significantly improve the temporal performance, approximately 24\% to 42\% in RMSE. It indicates that physics-informed losses can better supervise the temporal consistency for smooth solution changes. 

\paragraph{\textbf{Key modules for spatiotemporal matching}}
 There are two key components in our proposed \net\!\!: spatiotemporal attention (ST Attn) and feature split and reconstruction (FSR). To illustrate their effects, we use the Bahamas dataset to conduct the ablation studies and report the RMSE, MAE, and SSIM in Table~\ref{tab:module}. As introduced in Figure~\ref{fig:network}, the feature extraction module can be any image feature encoders. We choose RDN~\cite{rdn}, RCAN~\cite{rcan}, and SwinIR~\cite{swinir} as candidates for the feature extraction. Then, we add spatiotemporal attention and/or feature split and reconstruction (FSR) to see if they can positively affect the downscaling performance. We can see that 1) using SwinIR and RCAN is better than RDN as the feature encoder, as they can provide deeper feature representation, and rows 4 and 6 also prove that combining RCAN and spatiotemporal attention and FSR can provide the best performances. 2) Comparing columns 1\&4 and 2\&5, we can see that spatiotemporal attention can significantly improve the RMSE by about 0.021. Different feature encoders only have 0.08 improvements in RMSE and 0.002 in MAE. Given that SwinIR is a complex and time-consuming model, we choose RCAN, which balances downscaling quality and computation efficiency well.

\begin{table}[!th]\centering
\caption{\small \textbf{Ablation analysis on the key modules.} We report the results on the test datasets of the Bahamas using different combinations of modules. ST Attn refers to Spatiotemporal attention operation.} 
\renewcommand\arraystretch{1.4}
\resizebox{\columnwidth}{!}{
\begin{tabular}{c|c|ccccccccc}
\toprule
\multirow{3}{*}{Module}     & Feat encoder & RDN   & RCAN  & SwinIR & RDN   & RCAN  & SwinIR & \cellcolor{mistyrose}{RDN}   & \cellcolor{mistyrose}{RCAN}  & \cellcolor{mistyrose}{SwinIR} \\
                            & ST Attn      & \xmark     & \xmark     & \xmark      &  \cmark    & \cmark     & \cmark      & \cellcolor{mistyrose}{\cmark}     & \cellcolor{mistyrose}{\cmark}     & \cellcolor{mistyrose}{\cmark}      \\
                            & FSR          & \xmark     & \xmark     & \xmark      & \xmark     & \xmark     & \xmark      & \cellcolor{mistyrose}{\cmark}     & \cellcolor{mistyrose}{\cmark}     & \cellcolor{mistyrose}{\cmark}      \\
\multirow{3}{*}{Evaluation} & RMSE         & 0.054 & 0.050 & 0.035  & 0.033 & 0.031 & 0.025  & \cellcolor{mistyrose}{0.030} & \cellcolor{mistyrose}{0.028} & \cellcolor{mistyrose}{0.024}  \\
                            & MAE          & 0.023 & 0.021 & 0.016  & 0.013 & 0.013 & 0.011  & \cellcolor{mistyrose}{0.012} & \cellcolor{mistyrose}{0.011} & \cellcolor{mistyrose}{0.010}  \\
                            & SSIM         & 0.971 & 0.972 & 0.981  & 0.980 & 0.981 & 0.982  & \cellcolor{mistyrose}{0.982} & \cellcolor{mistyrose}{0.981} & \cellcolor{mistyrose}{0.982}  \\ \bottomrule
\end{tabular}%
}
\label{tab:module}
\end{table}

\begin{figure}[!th]\centering
 \includegraphics[width=\textwidth]{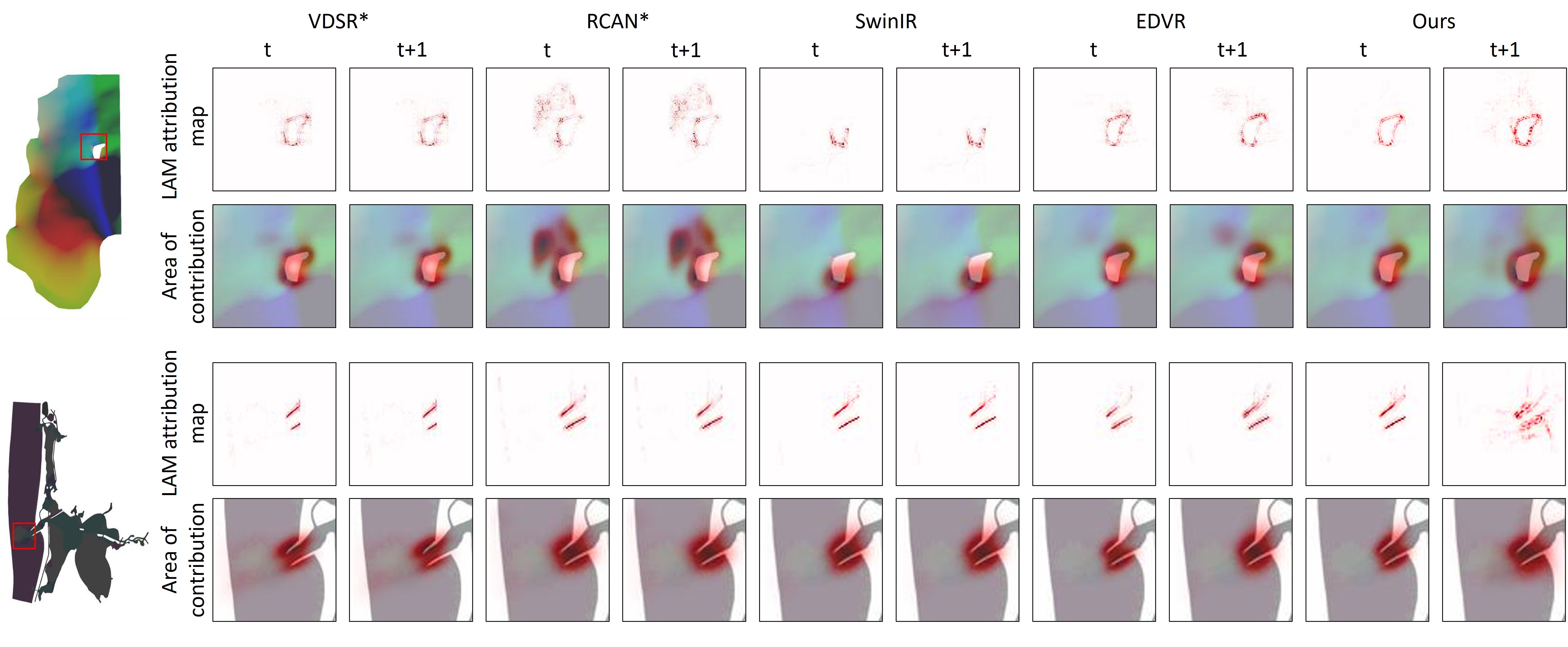}
 \caption{\small \textbf{Interpretation of downscaling method using attribute maps.}  The LAM maps represent the importance of each pixel in the input LR image w.r.t. the downscaled results of the patch marked with a red box. We also show the area of contribution to highlight the ROI region for computing feature correlations.}
 \label{fig:lam_compare}
\end{figure}

\begin{figure}[!thb]\centering
 \includegraphics[width=0.92\textwidth]{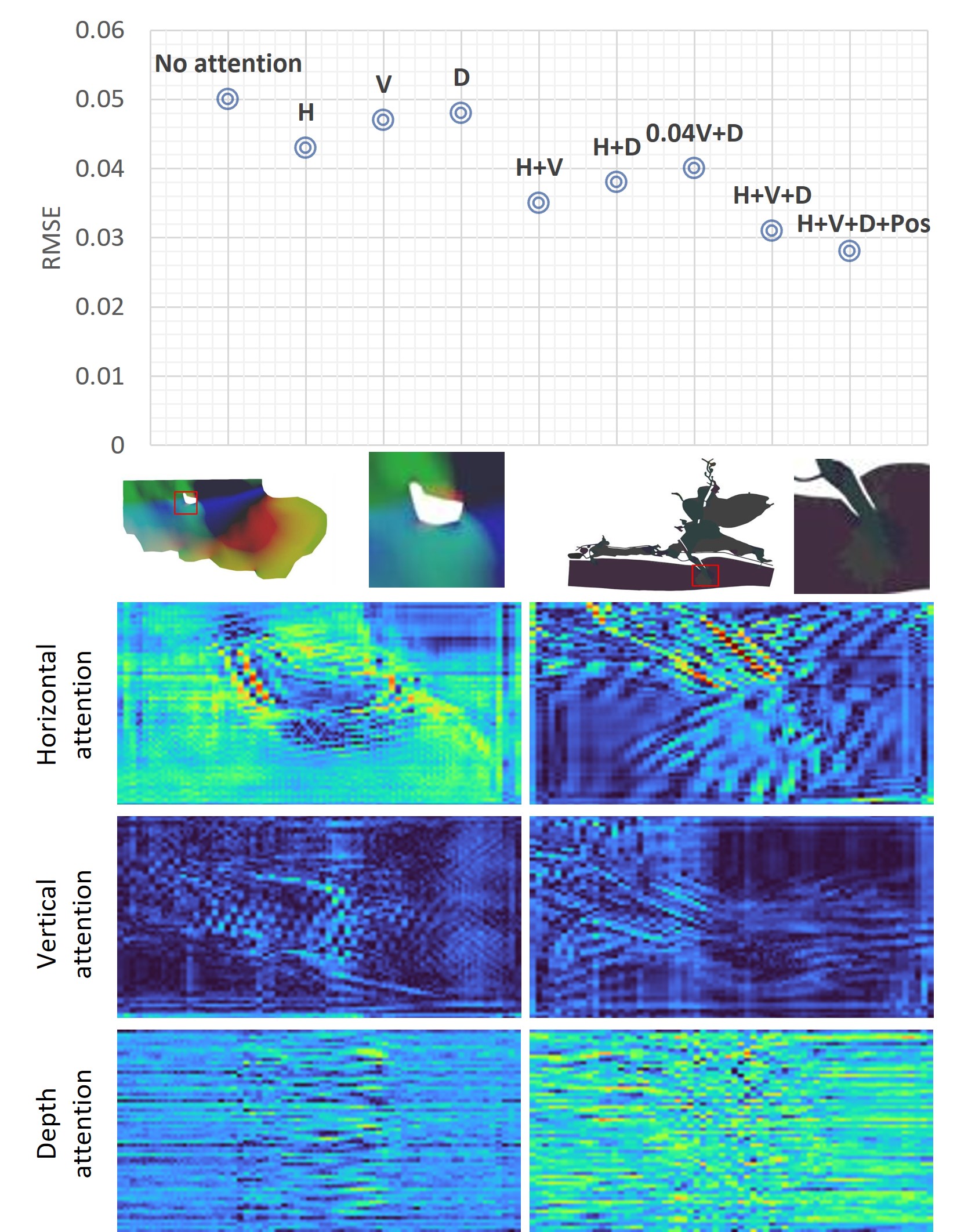}
 \caption{\small{\textbf{Visualization of spatiotemporal attention at different dimensions.} On top, we report the RMSE results of using different horizontal, vertical, and depth attention combinations. After applying horizontal, vertical, and depth attention, we show the feature map at the bottom.}}
 \label{fig:attn_compare}
\end{figure}

\paragraph{\textbf{Spatiotemporal attention}}
The key module of our proposed \net is the spatiotemporal attention. To visually understand its ability to capture the pixel information for simulation, we use Local Attribution Maps (LAM)~\cite{lam} to visualize the influences of neighborhood pixels on the region of interest (ROI) in Figure~\ref{fig:lam_compare}. The idea is to calculate the path-integrated gradient along a gradually changing path from the downscaling result to the LR input. In Figure~\ref{fig:lam_compare}, we show the contribution's pixel attribute map and region in both $t_n$ and $t_{n+1}$ LR inputs. The red box region is calculated and enlarged for comparison. We can see that EDVR and ours can explore wider regions, which means that they can extract a wider range of correlated features for calculation. RCAN can also explore larger regions because of its channel attention computation, however, it fails to show the differences caused by the temporal changes. On the contrary, we can observe that ours shows attribution changes (the red color ranges become larger from $t_n$ to $t_{n+1}$ LR image) because ours can explore the temporal changes and reflect the attention score on the different LR images.

\begin{figure}[!th]\centering
 \includegraphics[width=\textwidth]{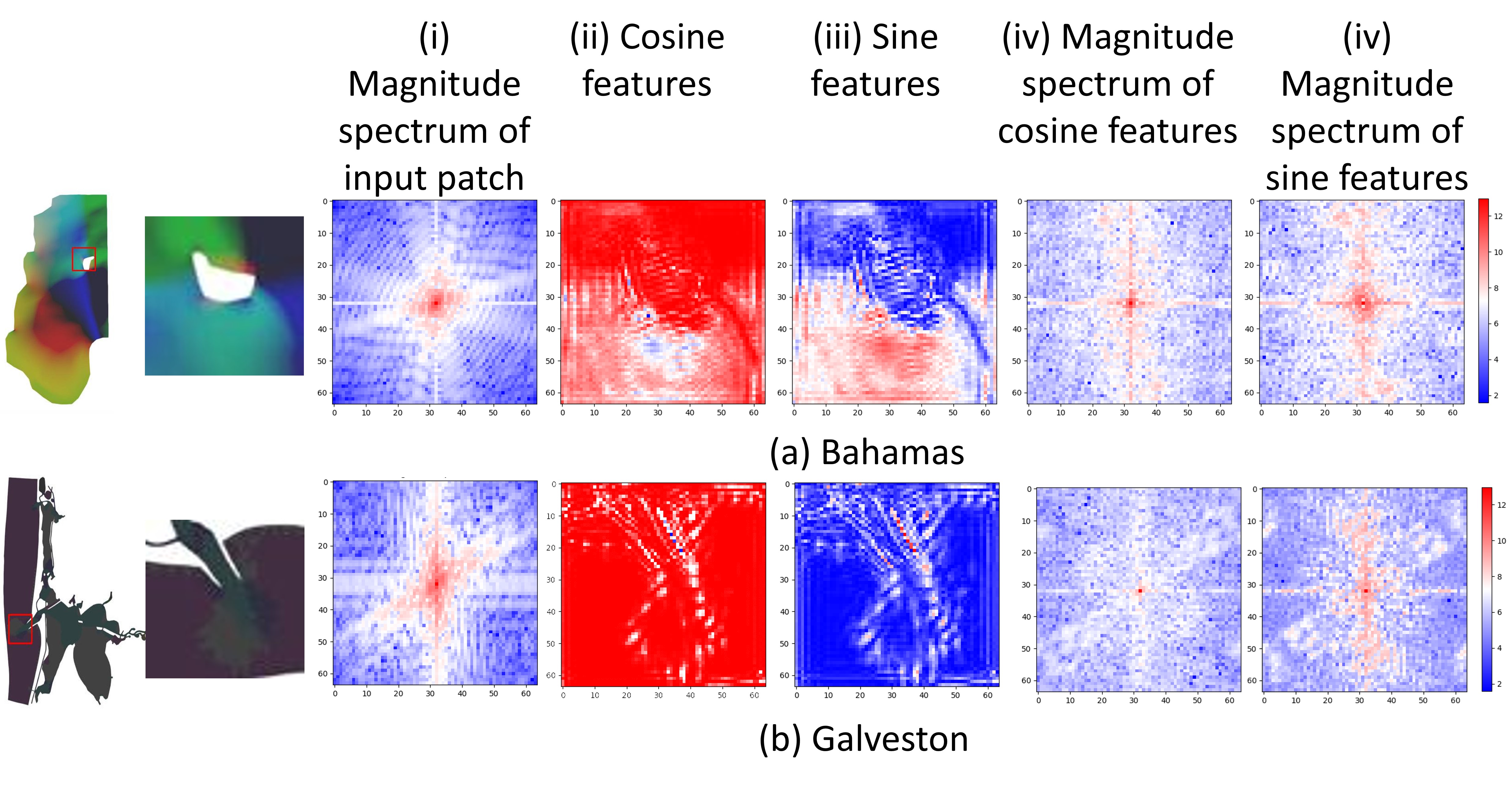}
 \caption{\small \textbf{Visualization of feature split and reconstruction.} We show the heatmap of the magnitude spectrum after applying FFT to the feature maps and the input LR image.}
 \label{fig:fsr}
\end{figure}

\begin{figure}[!th]\centering
 \includegraphics[width=\textwidth]{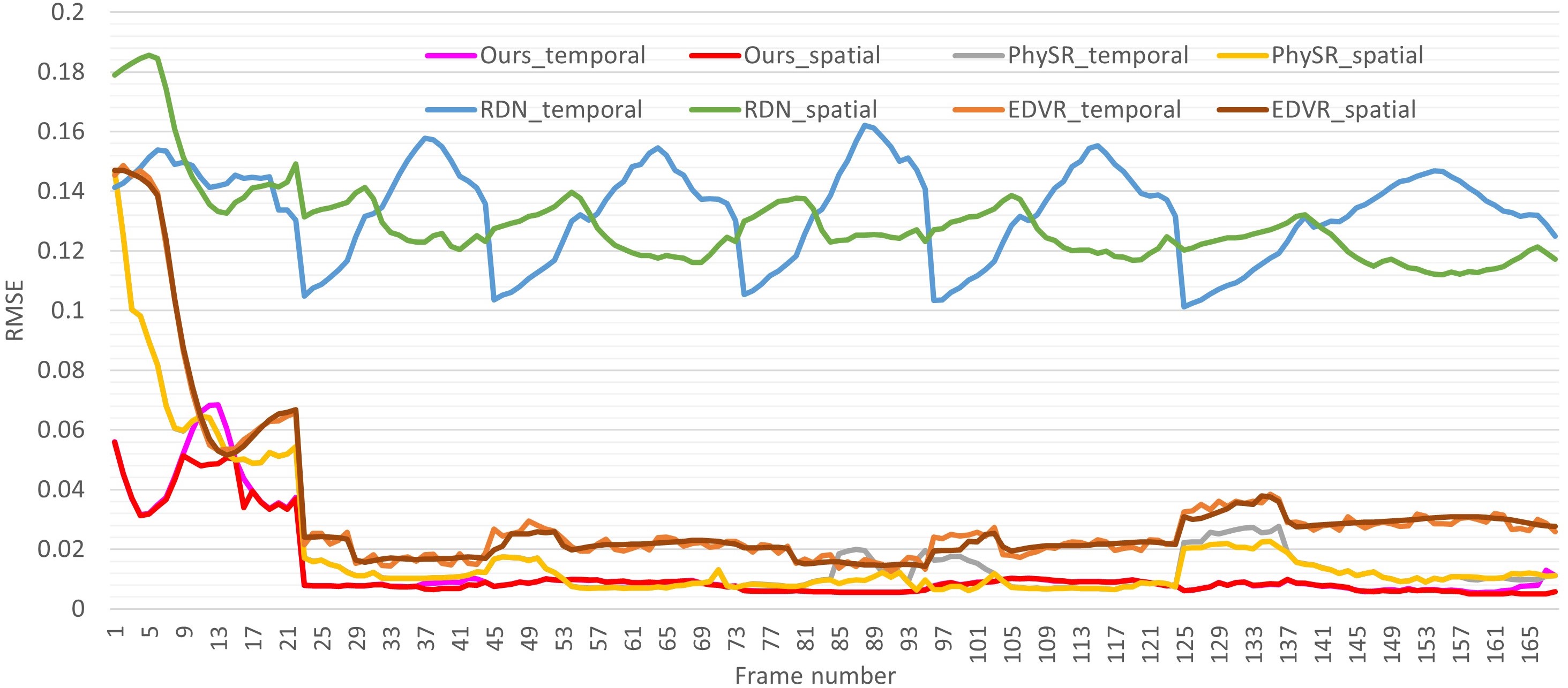}
 \caption{\small \textbf{Plot of frame-by-frame RMSE scores by using different methods.} It compares RMSE across 165 frames for different methods, evaluating spatial and temporal consistency. Temporal RMSE curves for RDN, EDVR and PhySR show prominent oscillations and higher overall error, indicating instability and poor motion coherence across frames. In contrast, spatial RMSE remains relatively flat, suggesting more consistent intra-frame quality. Our method (in pink for temporal and red for spatial) shows rapid convergence within the first 20 frames and maintains the lowest and most stable RMSE afterward, especially in the temporal domain, demonstrating superior consistency over time.}
 \label{fig:motion}
\end{figure}

Meanwhile, as introduced in Section III, the proposed spatiotemporal attention explores the horizontal(H)-, vertical(V)- and depth(D)-dimensional feature extraction hence, we conduct the experiments to compare the combination of three sub-attention modules and report the results in Figure~\ref{fig:attn_compare}. On top, we show the RMSE values of using attention operation at different dimensions. ``\textit{No attention}'' means no attention is used in the network.  ``\textit{H+V+D+Pos}'' means our final model with the positional encoded spatiotemporal attention module. We can conclude that 1) combining horizontal, vertical, and depth attention achieves the best performance, and 2) the feature maps reveal how the module focuses on the most important pixels in each direction. Specifically, by visualizing the attention maps, we can see that the model assigns higher weights to pixels corresponding to dominant features along horizontal, vertical, or depth directions.

\paragraph{\textbf{Feature split and reconstruction}}
For reconstruction, we are inspired by the recent development of image processing in the frequency domain~\cite{lte}. We apply the Fast Fourier Transform (FFT) to the LR image and the feature maps to see if they learn different aspects of feature representation. As shown in Figure~\ref{fig:fsr}, we show the feature map after cosine and sine operations. The magnitude spectrum maps show that they learn complementary information to predict the downscaled result. For instance, the cosine feature map learns the global low-frequency information, and the spectrum energy concentrates in the center. The sine feature map learns more about edge information, and the spectrum energy focuses on the vertical line.

\paragraph{\textbf{Motion consistency}}
We calculate the RMSE score of the Bahamas dataset to measure the motion consistency. We report intra-frame and inter-frame downscaling results to see if the RMSE score changes due to the spatial and temporal upsampling. We do not use optical flow or other motion estimation models because they are built on key feature point tracking. The figure illustrates RMSE trends over 165 frames for different super-resolution models. Our method (magenta for temporal, red for spatial) exhibits nearly identical curves in both spatial and temporal cases, indicating high motion consistency and robust reconstruction across frames. In contrast, RDN (blue and green) shows a large discrepancy between spatial and temporal RMSE, with noticeable oscillations in the temporal curve, revealing poor temporal coherence. EDVR and PhySR demonstrate smaller spatial-temporal mismatches (the gray and yellow curves of PhySR and light and dark brown curves of EDVR), but both produce higher RMSE overall compared to ours. These results highlight that our method achieves the best balance between spatial fidelity and temporal stability, making it particularly suitable for dynamic scenes like coastal simulations where motion preservation is critical.

\subsection{Downscaling adaptation to changing forcing data}\label{sec:Galveston_new_forcing}

\begin{figure}[!th]
\centering
\includegraphics[width=1\textwidth]{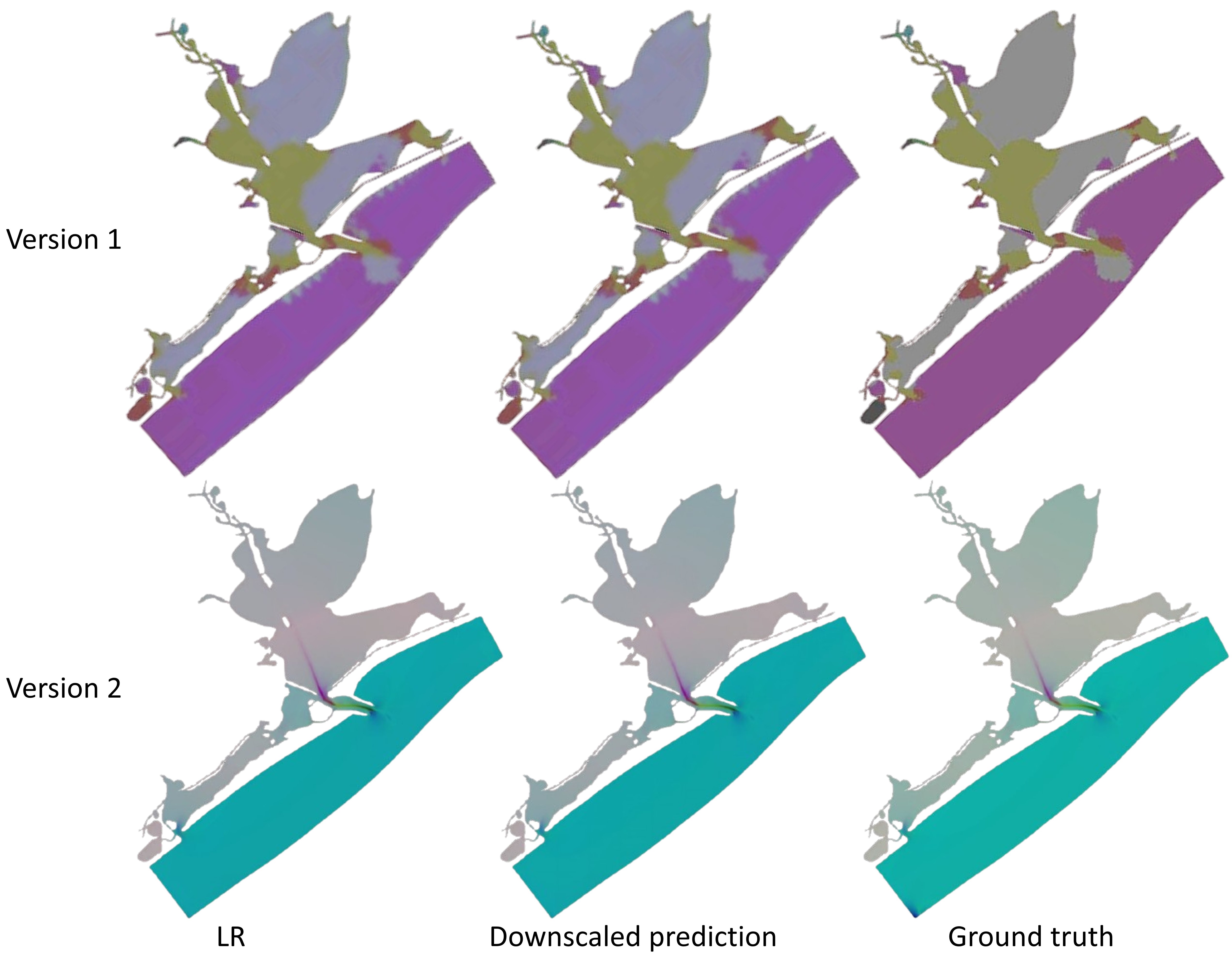}
\caption{\small Visual comparison (RGB values) of downscaling performance on Galveston datasets. Given the original simulation (version 1), and the modified tidal forcings (version 2), the proposed model provides consistent downscaling visual quality.}
    \label{fig:galveston_comp}
\end{figure}

We note that the Galveston data may also include other tidal forcings, such as wind and river inflow. To evaluate the adaptability of our model, we simulate an additional Galveston dataset with modified tidal forcings and directly apply the pretrained model to this new dataset without further fine-tuning. As shown in Figure~\ref{fig:galveston_comp}, the model exhibits similar performance across the two Galveston simulations, demonstrating that the proposed approach is simulation-agnostic. Table~\ref{tab:galveston_comp} further reports the quantitative downscaling results for both simulations, where only minor differences are observed across all evaluation metrics.

\begin{table}[!th]
\resizebox{\columnwidth}{!}{
\begin{tabular}{cccc|cccc}
\toprule
Metric & Resolution & version 1 & version 2 & Resolution & version 1 & version 2 \\ \midrule
RMSE & \multirow{5}{*}{3397$\times$13588} & 0.039 & 0.041 & \multirow{5}{*}{13588$\times$54352} & 0.062 & 0.066 \\
MAE & & 0.002 & 0.003 &  & 0.016 & 0.014 \\
SSIM & & 0.972 & 0.970 & & 0.979 & 0.975 \\
GMSD & & 0.043 & 0.045 &  & 0.079 & 0.079 \\
LPIPS & & 0.108 & 0.109 &  & 0.079 & 0.077 \\ 
 \bottomrule
\end{tabular}
}
\caption{Quantitative comparison of two Galveston simulations: version 1 with original simulation settings and version 2 with modified tidal forcings. The metrics are calculated on the average performance on the testing dataset without finetuning.}
\label{tab:galveston_comp}
\end{table}

\subsection{Storm surge during Hurricane Ike} \label{sec:realistic_flood} 

To illustrate the use of our proposed method to a realistic flood scenario, we generate data using the Discontinuous Galerkin Shallow Water Equation Model (DG-SWEM)~\cite{kubatko2006hp,Wichitrnithed2024}. This numerical model numerically approximates the SWE in \eqref{EQ:swe}, where the right hand side has been augmented with tidal forces, winds, and air pressure gradients. In addition, DG-SWEM has an advanced wetting and drying scheme to track the critical wet-dry interface during inundation~\cite{bunya2009wetting}. The selected event to simulate is Hurricane Ike (2008)~\cite{brown2010atlantic,hope2013hindcast}, which led to historic storm surge levels on the Texas coast. 

\begin{figure}[!th]\centering
 \includegraphics[width=0.8\textwidth]{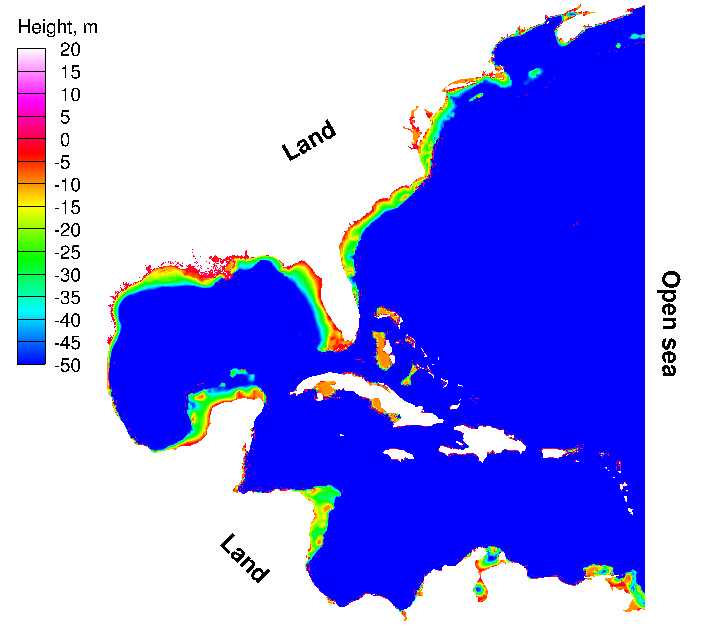}
 \caption{\small{Computational domain for Ike storm surge simulations, bathymetry cut off at 50m below and 20m above sea level according to the North Atlantic Vertical Datum of 1988 (in meters).}}
 \label{fig:30m}
\end{figure}

With DG-SWEM, we simulate storm surge during this hurricane using winds and air pressures from high-resolution reanalysis data as described in~\cite{hope2013hindcast}, and tidal forcings from the tidal constituents (M2, N2, O1, S2, K1, P1, Q1, K2) obtained from the TPXO9 model~\cite{egbert2002efficient}. We run DG-SWEM using a Local Lax-Friedrichs flux,  polynomial order one, and a three stage - second order explicit Runge Kutta time stepping scheme.
The computational domain in shown in Figure~\ref{fig:30m} and covers the entire ocean to the west of the 60$^\circ$ meridian. The simulations cover the time period of September 5, 2008 - September 14, 2008 (noon to noon), with a time step size of $0.5$ seconds. Two meshes were used for simulation -- see Table~\ref{tab:ike-data} for details of the mesh resolutions and computational resources used in the simulation on the Frontera cluster at the Texas Advanced Computing Center (TACC). The two meshes and corresponding simulations differ only in mesh resolution, see~\cite{contreras2023channel} for further detail and validation of these datasets.

\begin{table}[!th]\centering
\caption{\small \textbf{Summary of the data used in Ike storm surge simulations.} The ``Elem.'' value represents the number of mesh elements, `Min/Max' column contains the minimum/maximum element size in meters. In the ``\# CPU cores`` column, we list the total number of Intel Xeon Platinum 8280 `Cascade Lake' CPU cores (56 cores per node of Frontera cluster) used in the simulation and `Wallclock time' is the corresponding total wallclock time of the simulation.}
\begin{tabular}{c|ccc}
\toprule 
Elem. & Min/Max (meters) & \# CPU cores & Wallclock time  \\ 
\midrule
3,832,707  & 30/24000  & 2128   & 00:53:05  \\
15,459,336  &  120/24000   & 2128   & 11:24:41  \\ 
\bottomrule
\end{tabular}
\label{tab:ike-data}
\end{table}

\begin{figure}[!th]\centering
\includegraphics[width=\textwidth]{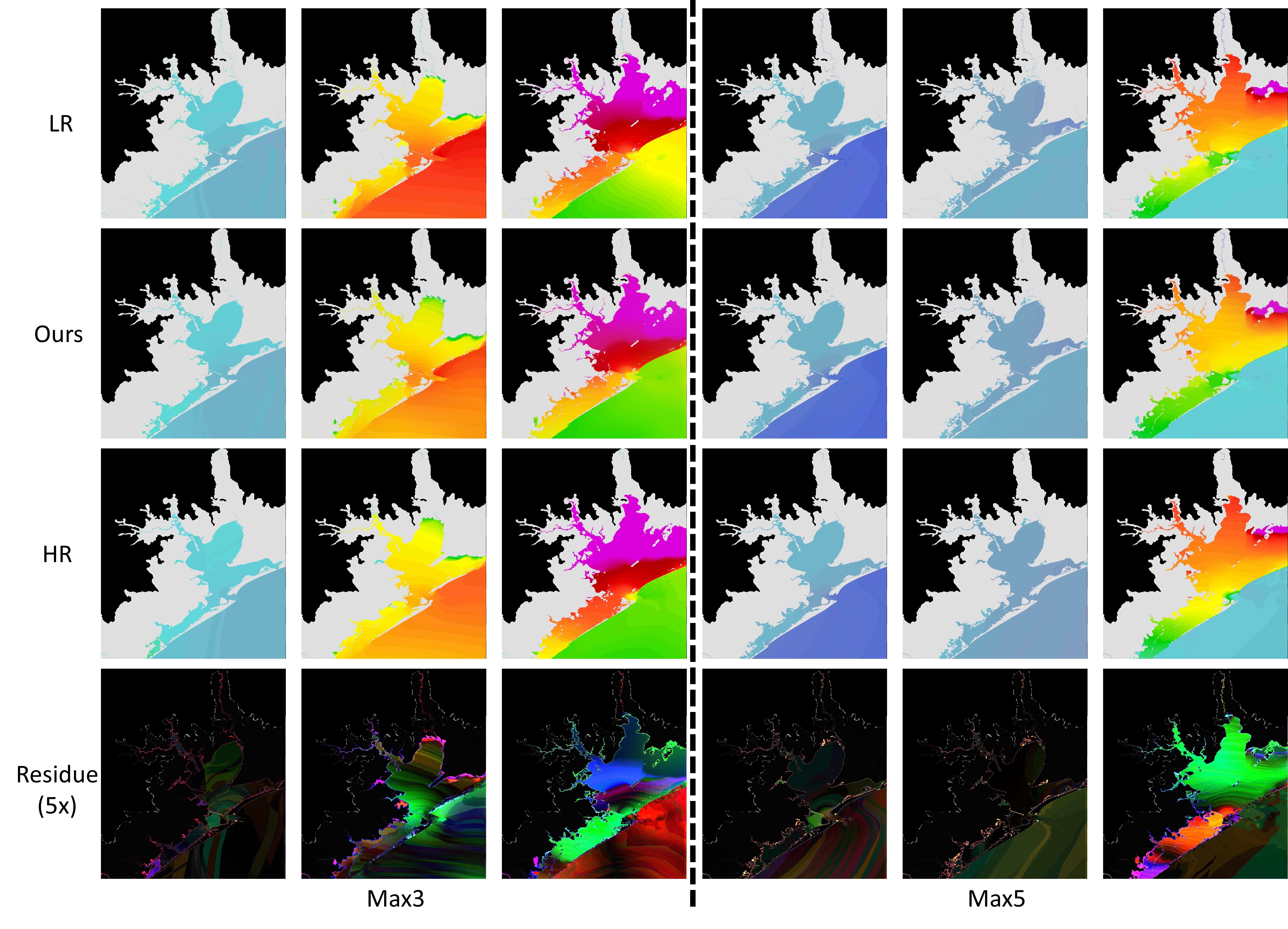}
 \caption{\small \textbf{Visual illustration of the proposed method for spatio-temporal downscaling on the flooding scenario.} Given input simulated flooding data, we apply our model for 2$\times$ temporal interpolation and 4$\times$. To highlight the differences, we compute the residuals between the model predictions and the target high-resolution (HR) data, amplify the residuals by a factor of 5, and visualize them as RGB images in the last row.}
 \label{fig:flood}
\end{figure}

\begin{table}[!th]\centering
\caption{\small \textbf{Quantitative results on real-world downscaling.} The evaluation is conducted on keyframes and interframes, where keyframes denote time-aligned LR–HR frame pairs, while interframes refer to intermediate time steps predicted by the neural network using two adjacent LR frames.} 
\renewcommand\arraystretch{1.4}
\resizebox{\columnwidth}{!}{
\begin{tabular}{c|cccc|cccc}
\toprule
\multirow{2}{*}{Frames} & \multicolumn{4}{c|}{Keyframe} & \multicolumn{4}{c}{Interframes} \\
 & \multicolumn{2}{c}{LR} & \multicolumn{2}{c|}{Ours} & \multicolumn{2}{c}{Bicubic} & \multicolumn{2}{c}{Ours} \\ \midrule
\multirow{2}{*}{Metric} & MAE & MSE & MAE & MSE & MAE & MSE & MAE & MSE \\
 & 0.0198 & 0.0052 & 0.0133 & 0.00251 & 0.0256 & 0.0057 & 0.0197 & 0.00295 \\ \bottomrule
\end{tabular}%
}
\label{tab:real_world}
\end{table}

To validate our method's adaptation to a real-world flooding scenario, we generate a new dataset based on Hurricane Ike. The data are converted into video sequences of 205 frames at a resolution of $5803\times 5906$. We randomly split the data into training (165) and test (40) datasets. We use the training data to fine-tune our pretrained model for 100 epochs (approximately 2 hours). We then evaluate the model's performance on the test dataset using the most recent model checkpoint. Figure~\ref{fig:flood} presents a visual comparison between the low-resolution (LR) inputs, our spatio-temporally downscaled results, and the high-resolution (HR) reference data under a real-world flooding scenario. Compared with the LR inputs, our method substantially enhances spatial details. It produces smoother, more coherent inundation structures that are visually consistent with the HR observations, particularly along the main flood fronts and large-scale flow patterns. The overall color gradients and shoreline geometries reconstructed by our model closely resemble those of the HR data, indicating effective recovery of fine-scale spatial variations during both moderate and extreme flooding stages. This observation is further supported by the residual visualizations (last row), in which the amplified errors remain confined mainly to narrow boundary regions and localized high-gradient areas, whereas the majority of flooded regions exhibit low residual magnitudes. In contrast, the LR inputs fail to capture these sharp transitions and localized structures. Although minor discrepancies persist in regions with complex topography or rapidly varying flood dynamics, the overall residual magnitude remains limited, suggesting that our method preserves both global consistency and local fidelity. These results demonstrate the strong capability of the proposed model to bridge the resolution gap between simulated inputs and HR flooding patterns, even in challenging real-world conditions. Quantitative comparison is shown in Table~\ref{tab:real_world}. We separately report the results of spatial downscaling (keyframes) and temporal downscaling (interframes). For spatial downscaling, paired LR and HR data are available, enabling direct comparison of the HR ground truth with the model predictions. For temporal downscaling, the model takes neighboring LR frames as input to estimate the intermediate frames via temporal interpolation. For comparison, we use temporal bicubic as the baseline method. As shown in the table, our method achieves improvements of approximately 0.002 in MSE and 0.006 in MAE relative to the baseline.

\subsection{Limitations and further discussion}
A key limitation of the proposed framework is that optimal performance on previously unseen coastal regions generally requires fine-tuning with local data. This behavior arises from the strong domain-specific physical conditions inherent to different coastal environments, including shoreline geometry, bathymetry, and region-dependent hydrodynamic patterns, all of which influence fine-scale water movements. In practice, we identify two deployment modes. First, the pretrained model can be directly applied to a new region in a zero-shot manner when the land-sea mask is available. While this approach enables rapid, low-cost testing, its performance may vary depending on the degree of similarity between the source and target regions. Second, for operational use, fine-tuning the pretrained model with historical coarse--fine data pairs from the target region significantly improves accuracy by adapting the model to the local data distribution. Importantly, because the model is pretrained on physics-based simulations, the computational overhead of fine-tuning is modest, making this strategy feasible in real-world applications (see Section~5.5).

From a modeling perspective, our current design treats landmass as a hard constraint by assigning zero values over land, which effectively enforces physical boundaries but may limit the representation of complex coastal interactions near shorelines. A promising direction for future work is to reformulate coastal downscaling on irregular meshes as a graph-based learning problem. By representing coastal simulations as graphs and employing graph neural networks, water movement could propagate along vertices and edges, enabling the model to capture nonlocal dependencies and complex boundary interactions more naturally. Such an approach may further enhance domain transferability and robustness across diverse coastal regions.

\section{Conclusion}

In this paper, we propose \net\!\!\!, a spatiotemporal downscaling neural network for efficient coastal ocean simulation. The proposed spatiotemporal attention fully utilizes the neighborhood pixels across space and time via pixel shuffle to estimate the feature correlations. It can explicitly model the U-, V- and $\xi$-view changes over time as spatial feature correlations. The grid location is also embedded as the positional code for axis-aware attention estimation. Finally, we implement the novel feature split and reconstruction to project the features into the frequency domain for final enhancement. The initial spatiotemporal bilinear operator is used to encourage the network to learn the residues between LR and HR data. The proposed physics-informed losses also help to optimize the neural network with better quantitative results. In order to train the neural network, we propose a novel coastal simulation dataset on the Bahamas and Galveston. We use it for model training and evaluation. From extensive experiments and comparisons with other state-of-the-art methods, we can conclude that ours can achieve superior downscaling quality with fast computation. In addition, we demonstrate the transferability of our methodology to other datasets and scenarios without extensive training, including flooding simulations not covered in our original training data.

Because the proposed methodology guarantees the mass conservation for arbitrarily long simulation times, it is particularly attractive for ocean, climate, and weather models for which high resolution simulations are desirable but computationally very expensive. In addition, it paves a novel direction for neural downscaling models for multigrid simulation, and we are interested in further exploration of ultra-resolution reconstruction and continuous downscaling.

\section*{Software and Data Availability}
Software name: DNNCS | Developer: Zhi-Song Liu | First year available: 2025 | Hardware requirements: PC with Nvidia GPU | Software requirements: Pytorch, Pytorch Lightning, OpenCV, einops | Program language: Pytorch | Availability: \url{https://github.com/Holmes-Alan/DNNCS} | License: MIT | Test data and pretrained model: Same as the code repository

\section*{Acknowledgments}
A.\ Rupp has been supported by the Academy of Finland's grants number 350101, 354489, 359633, 358944, Business Finland's project number 539/31/2023, and Deutsche Forschungsgemeinschaft (DFG, German Research Foundation) grant number 577175348. Markus Büttner was partially supported by DFG under grant AI 117/7-1 within the project 502500606 `Performance-optimized co-design of ocean modeling software on FPGAs'.

Bernhard Kainz acknowledges the HPC resources provided by NHR@FAU under the NHR projects b143dc and b180dc. NHR funding is provided by federal and Bavarian state authorities. NHR@FAU hardware is partially funded by the DFG - 440719683. Additional support was  received by the ERC-project MIA-NORMAL 101083647,  DFG 513220538, 512819079, and by the state of Bavaria (HTA).

\printbibliography

@article{Bui16,
author = {Bui-Thanh, T.},
title = {{Construction and Analysis of {HDG} Methods for Linearized Shallow Water Equations}},
journal = {SIAM Journal on Scientific Computing},
volume = {38},
number = {6},
pages = {A3696-A3719},
year = {2016},
doi = {10.1137/16M1057243},
}

@article{MorauRB23,
    author = {Moraru, Adina and Rüther, Nils and Bruland, Oddbjørn},
    title = "{Investigating optimal 2D hydrodynamic modeling of a recent flash flood in a steep Norwegian river using high-performance computing}",
    journal = {Journal of Hydroinformatics},
    volume = {25},
    number = {5},
    pages = {1690-1712},
    year = {2023},
    month = {09},
    issn = {1464-7141},
    doi = {10.2166/hydro.2023.012},
}

@article{ButtingerKHCSBW22,
title = {An integrated {GPU}-accelerated modeling framework for high-resolution simulations of rural and urban flash floods},
journal = {Environmental Modelling \& Software},
volume = {156},
pages = {105480},
year = {2022},
issn = {1364-8152},
doi = {https://doi.org/10.1016/j.envsoft.2022.105480},
author = {Andreas Buttinger-Kreuzhuber and Artem Konev and Zsolt Horváth and Daniel Cornel and Ingo Schwerdorf and Günter Blöschl and Jürgen Waser},
}

@article{MoralesEtAl20,
    author = {Morales-Hernández, M. and Sharif, M. B. and Gangrade, S. and Dullo, T. T. and Kao, S.-C. and Kalyanapu, A. and Ghafoor, S. K. and Evans, K. J. and Madadi-Kandjani, E. and Hodges, B. R.},
    title = "{High-performance computing in water resources hydrodynamics}",
    journal = {Journal of Hydroinformatics},
    volume = {22},
    number = {5},
    pages = {1217-1235},
    year = {2020},
    month = {03},
    issn = {1464-7141},
    doi = {10.2166/hydro.2020.163},
}

@article{SamiiKMD19,
author = {Samii, A. and Kazhyken, K. and Michoski, C.E. and Dawson, C.},
year = {2019},
month = {09},
title = {A Comparison of the Explicit and Implicit Hybridizable Discontinuous {G}alerkin Methods for Nonlinear Shallow Water Equations},
volume = {80},
journal = {Journal of Scientific Computing},
doi = {10.1007/s10915-019-01007-z}
}

@article{AizingerDawson2002,
  Title                    = {A discontinuous {G}alerkin method for two-dimensional flow and transport in shallow water},
  Author                   = {V. Aizinger and C. Dawson},
  Journal                  = {Advances in Water Resources},
  Year                     = {2002},
  Number                   = {1},
  Pages                    = {67-84},
  Volume                   = {25},
  Doi                      = {10.1016/S0309-1708(01)00019-7}
}

@article{CockburnShuRKDG21989,
author = {Cockburn, B. and Shu, C.-W.},
title = {{TVB} {Runge-Kutta} local projection discontinuous {Galerkin} finite element method for conservation laws. {II. General framework}},
journal = {Math. Comp.},
volume = {52},
pages = {411-435},
year = {1989},
doi = {10.1090/S0025-5718-1989-0983311-4}
}

@InProceedings{Tsunawi2011,
author="Androsov, Alexey
and Behrens, J{\"o}rn
and Danilov, Sergey",
editor="Krause, Egon
and Shokin, Yurii
and Resch, Michael
and Kr{\"o}ner, Dietmar
and Shokina, Nina",
title="Tsunami Modelling with Unstructured Grids. Interaction between Tides and Tsunami Waves",
booktitle="Computational Science and High Performance Computing IV",
year="2011",
publisher="Springer Berlin Heidelberg",
address="Berlin, Heidelberg",
pages="191--206",
}

@Article{Baptista2011,
AUTHOR = {Baptista, M. A. and Miranda, J. M. and Omira, R. and Antunes, C.},
TITLE = {Potential inundation of {L}isbon downtown by a 1755-like tsunami},
JOURNAL = {Natural Hazards and Earth System Sciences},
VOLUME = {11},
YEAR = {2011},
NUMBER = {12},
PAGES = {3319--3326},
DOI = {10.5194/nhess-11-3319-2011}
}

@article{HajdukHAR2018,
Author = {H. Hajduk and B. R. Hodges and V. Aizinger and B. Reuter},
Title = {Locally Filtered Transport for computational efficiency in multi-component advection-reaction models},
Journal = {Environmental Modelling \& Software},
Year = {2018},
Volume = {102},
Pages = {185-198},
Doi = {10.1016/j.envsoft.2018.01.003},
}

@article{HauckAFHR2020,
         author = {M. Hauck and V. Aizinger and F. Frank and H. Hajduk and A. Rupp},
           title = {Enriched {G}alerkin method for the shallow-water equations},
           volume = {11},
          note = {Article number 31},
           year = {2020},
          journal = {GEM : International Journal on Geomathematics},
          doi = {10.1007/s13137-020-00167-7},
}

@article{FaghihNainiKAZGK2020,
title = "Quadrature-free discontinuous {G}alerkin method with code generation features for shallow water equations on automatically generated block-structured meshes",
journal = "Advances in Water Resources",
volume = "138",
pages = "103552",
year = "2020",
issn = "0309-1708",
author = "S. Faghih-Naini and S. Kuckuk and V. Aizinger and D. Zint and R. Grosso and H. K{\"o}stler",
}

@inproceedings{KenterSFA2021,
author = {Kenter, Tobias and Shambhu, Adesh and Faghih-Naini, Sara and Aizinger, Vadym},
title = {{Algorithm-Hardware Co-Design of a Discontinuous Galerkin Shallow-Water Model for a Dataflow Architecture on FPGA}},
year = {2021},
isbn = {9781450385633},
booktitle = {Proceedings of the Platform for Advanced Scientific Computing Conference},
articleno = {13},
numpages = {11},
location = {Geneva, Switzerland},
series = {PASC '21}
}

@article{FaghihNainiKZKKA2023,
title = {Discontinuous {G}alerkin method for the shallow water equations on complex domains using masked block-structured grids},
journal = {Advances in Water Resources},
pages = {104584},
year = {2023},
issn = {0309-1708},
doi = {10.1016/j.advwatres.2023.104584},
author = {Sara Faghih-Naini and Sebastian Kuckuk and Daniel Zint and Samuel Kemmler and Harald Köstler and Vadym Aizinger},
}

@inproceedings{FajKFPA2023,
author = {Faj, Jennifer and Kenter, Tobias and Faghih-Naini, Sara and Plessl, Christian and Aizinger, Vadym},
title = {{Scalable Multi-FPGA Design of a Discontinuous Galerkin Shallow-Water Model on Unstructured Meshes}},
year = {2023},
isbn = {9798400701900},
publisher = {Association for Computing Machinery},
address = {New York, NY, USA},
doi = {10.1145/3592979.3593407},
booktitle = {Proceedings of the Platform for Advanced Scientific Computing Conference},
articleno = {8},
numpages = {12},
location = {Davos, Switzerland},
series = {PASC '23}
}

@InProceedings{AltKFFOPAHK2023,
author={Alt, Christoph
and Kenter, Tobias
and Faghih-Naini, Sara
and Faj, Jennifer
and Opdenh{\"o}vel, Jan-Oliver
and Plessl, Christian
and Aizinger, Vadym
and et al.},
title={{Shallow Water DG Simulations on FPGAs: Design and Comparison of a Novel Code Generation Pipeline}},
booktitle={High Performance Computing},
year={2023},
pages={86-105},
}

@inproceedings{BuettnerAKKPA2024,
author = {B\"{u}ttner, Markus and Alt, Christoph and Kenter, Tobias and K\"{o}stler, Harald and Plessl, Christian and Aizinger, Vadym},
title = {{Enabling Performance Portability for Shallow Water Equations on CPUs, GPUs, and FPGAs with SYCL}},
year = {2024},
isbn = {9798400706394},
publisher = {Association for Computing Machinery},
address = {New York, NY, USA},
doi = {10.1145/3659914.3659925},
booktitle = {Proceedings of the Platform for Advanced Scientific Computing Conference},
articleno = {11},
numpages = {12},
location = {Zurich, Switzerland},
series = {PASC '24}
}

@article{AizingerPDPN2013,
Author = {V. Aizinger and J. Proft and C. Dawson and D. Pothina and S. Negusse},
Title = {{A three-dimensional discontinuous Galerkin model applied to the baroclinic simulation of Corpus Christi Bay}},
Journal = {Ocean Dynamics},
Year = {2013},
Volume = {63},
Number = {1},
Pages = {89-113},
Publisher = {Springer},
Doi = {10.1007/s10236-012-0579-8},
}

@article{DawsonAizinger2005,
Author = {C. Dawson and V. Aizinger},
Title = {A discontinuous {G}alerkin method for three-dimensional shallow water equations},
Journal = {Journal of Scientific Computing},
Year = {2005},
Volume = {22},
Number = {1-3},
Pages = {245-267},
Publisher = {Springer},
Doi = {10.1007/s10915-004-4139-3},
}

@article{BuettnerAKKPA2025,
            year = {2025},
         journal = {The Journal of Supercomputing},
           title = {Analyzing performance portability for a {SYCL} implementation of the {2D} shallow water equations},
          number = {6},
          volume = {81},
            issn = {1573-0484},
          author = {B{\"u}ttner, Markus and Alt, Christoph and Kenter, Tobias and K{\"o}stler, Harald and Plessl, Christian and Aizinger, Vadym},
            url = {https://doi.org/10.1007/s11227-025-07063-7},
 	   doi = {10.1007/s11227-025-07063-7},
}

@article{FaghihNainiAKAK2025,
          number = {1},
          volume = {16},
         journal = {GEM : International Journal on Geomathematics},
            year = {2025},
           title = {p-adaptive discontinuous {G}alerkin method for the shallow water equations on heterogeneous computing architectures},
          author = {Faghih-Naini, Sara and Aizinger, Vadym and Kuckuk, Sebastian and Angersbach, Richard and K{\"o}stler, Harald},
            issn = {1869-2672},
            url = {https://doi.org/10.1007/s13137-025-00267-2},
 	   doi = {10.1007/s13137-025-00267-2},
}

@article {ADCIRC2010,
      author = {S. Bunya and J. C. Dietrich and J. J. Westerink and B. A. Ebersole and et al.},
      title = {{A High-Resolution Coupled Riverine Flow, Tide, Wind, Wind Wave, and Storm Surge Model for Southern Louisiana and Mississippi. Part I: Model Development and Validation}},
      journal = {Monthly Weather Review},
      year = {2010},
      volume = {138},
      number = {2},
      pages= {345 - 377},
}

@article{Xing2010,
title = {Positivity-preserving high order well-balanced discontinuous {G}alerkin methods for the shallow water equations},
journal = {Advances in Water Resources},
volume = {33},
number = {12},
pages = {1476-1493},
year = {2010},
issn = {0309-1708},
doi = {https://doi.org/10.1016/j.advwatres.2010.08.005},
author = {Yulong Xing and Xiangxiong Zhang and Chi-Wang Shu},
}

@article{Bunya2009,
title = {A wetting and drying treatment for the Runge–Kutta discontinuous {G}alerkin solution to the shallow water equations},
journal = {Computer Methods in Applied Mechanics and Engineering},
volume = {198},
number = {17},
pages = {1548-1562},
year = {2009},
issn = {0045-7825},
author = {Shintaro Bunya and Ethan J. Kubatko and Joannes J. Westerink and Clint Dawson},
}

@article{Ern2008,
author = {Ern, A. and Piperno, S. and Djadel, K.},
title = {{A well-balanced Runge–Kutta discontinuous Galerkin method for the shallow-water equations with flooding and drying}},
journal = {International Journal for Numerical Methods in Fluids},
volume = {58},
number = {1},
pages = {1-25},
year = {2008}
}

@ARTICLE{Eskilsson2004,
	author = {Eskilsson, C. and Sherwin, Spencer J.},
	title = {{A triangular spectral/hp discontinuous Galerkin method for modelling 2D shallow water equations}},
	year = {2004},
	journal = {International Journal for Numerical Methods in Fluids},
	volume = {45},
	number = {6},
	pages = {605-623},
	doi = {10.1002/fld.709},
}

@article{Tumolo2015,
author = {Tumolo, Giovanni and Bonaventura, Luca},
title = {{A semi-implicit, semi-Lagrangian discontinuous Galerkin framework for adaptive numerical weather prediction}},
journal = {Quarterly Journal of the Royal Meteorological Society},
volume = {141},
number = {692},
pages = {2582-2601},
doi = {https://doi.org/10.1002/qj.2544},
year = {2015}
}

@Article{Shaw2021,
AUTHOR = {Shaw, J. and Kesserwani, G. and Neal, J. and Bates, P. and Sharifian, M. K.},
TITLE = {{LISFLOOD-FP 8.0: the new discontinuous Galerkin shallow-water solver for multi-core CPUs and GPUs}},
JOURNAL = {Geoscientific Model Development},
VOLUME = {14},
YEAR = {2021},
NUMBER = {6},
PAGES = {3577--3602},
DOI = {10.5194/gmd-14-3577-2021}
}

@article{Wichitrnithed2024,
title = {A discontinuous {G}alerkin finite element model for compound flood simulations},
journal = {Computer Methods in Applied Mechanics and Engineering},
volume = {420},
pages = {116707},
year = {2024},
issn = {0045-7825},
doi = {https://doi.org/10.1016/j.cma.2023.116707},
author = {Chayanon Wichitrnithed and Eirik Valseth and Ethan J. Kubatko and Younghun Kang and Mackenzie Hudson and Clint Dawson},
}

@ARTICLE{SSIM, 
author={Z. Wang and A. C. {Bovik} and H. R. {Sheikh} and E. P. {Simoncelli}}, 
journal={IEEE Transactions on Image Processing}, 
title={Image quality assessment: from error visibility to structural similarity}, 
year={2004}, 
volume={13}, 
number={4}, 
pages={600-612},  
ISSN={1057-7149}, 
month={April}
}

@inproceedings{LPIPS,
title={The Unreasonable Effectiveness of Deep Features as a Perceptual Metric},
author={Zhang, Richard and Isola, Phillip and Efros, Alexei A and Shechtman, Eli and Wang, Oliver},
booktitle={CVPR},
year={2018}
}

@ARTICLE{gmsd,
  author={Xue, Wufeng and Zhang, Lei and Mou, Xuanqin and Bovik, Alan C.},
  journal={IEEE Transactions on Image Processing}, 
  title={{Gradient Magnitude Similarity Deviation: A Highly Efficient Perceptual Image Quality Index}}, 
  year={2014},
  volume={23},
  number={2},
  pages={684-695},
}

@InProceedings{edvr,
          author = {Wang, Xintao and Chan, Kelvin C.K. and Yu, Ke and Dong, Chao and Loy, Chen Change},
          title = {{EDVR: Video Restoration with Enhanced Deformable Convolutional Networks}},
          booktitle = {The IEEE Conference on Computer Vision and Pattern Recognition (CVPR) Workshops},
          month = {June},
          year = {2019}
}

@INPROCEEDINGS{vdsr,
  author={Kim, Jiwon and Lee, Jung Kwon and Lee, Kyoung Mu},
  booktitle={2016 IEEE Conference on Computer Vision and Pattern Recognition (CVPR)}, 
  title={{Accurate Image Super-Resolution Using Very Deep Convolutional Networks}}, 
  year={2016},
  volume={},
  number={},
  pages={1646-1654},
  doi={10.1109/CVPR.2016.182}
}

@INPROCEEDINGS{rdn,
  author={Zhang, Yulun and Tian, Yapeng and Kong, Yu and Zhong, Bineng and Fu, Yun},
  booktitle={2018 IEEE/CVF Conference on Computer Vision and Pattern Recognition}, 
  title={{Residual Dense Network for Image Super-Resolution}}, 
  year={2018},
  volume={},
  number={},
  pages={2472-2481},
  doi={10.1109/CVPR.2018.00262}
}

@inproceedings{rcan,
    title={{Image Super-Resolution Using Very Deep Residual Channel Attention Networks}},
    author={Zhang, Yulun and Li, Kunpeng and Li, Kai and Wang, Lichen and Zhong, Bineng and Fu, Yun},
    booktitle={ECCV},
    year={2018}
}

@inproceedings{stablesr,
    author = {Wang, Jianyi and Yue, Zongsheng and Zhou, Shangchen and Chan, Kelvin CK and Loy, Chen Change},
    title = {{Exploiting Diffusion Prior for Real-World Image Super-Resolution}},
    booktitle = {International Journal of Computer Vision},
    year = {2024}
}

@INPROCEEDINGS{swinir,
  author={Liang, Jingyun and Cao, Jiezhang and Sun, Guolei and Zhang, Kai and Van Gool, Luc and Timofte, Radu},
  booktitle={2021 IEEE/CVF International Conference on Computer Vision Workshops (ICCVW)}, 
  title={{SwinIR: Image Restoration Using Swin Transformer}}, 
  year={2021},
  volume={},
  number={},
  pages={1833-1844},
  doi={10.1109/ICCVW54120.2021.00210}
}

@article{physr,
  title={{PhySR: Physics-informed Deep Super-resolution for Spatiotemporal Data}},
  author={Ren, Pu and Rao, Chengping and Liu, Yang and Ma, Zihan and Wang, Qi and Wang, Jian-Xun and Sun, Hao},
  journal={Journal of Computational Physics},
  pages={112438},
  year={2023},
  publisher={Elsevier}
}

@ARTICLE{srcnn,
  author={Dong, Chao and Loy, Chen Change and He, Kaiming and Tang, Xiaoou},
  journal={IEEE Transactions on Pattern Analysis and Machine Intelligence}, 
  title={{Image Super-Resolution Using Deep Convolutional Networks}}, 
  year={2016},
  volume={38},
  number={2},
  pages={295-307},
  doi={10.1109/TPAMI.2015.2439281}
}

@INPROCEEDINGS{sr_1,
  author={Liu, Zhi-Song and et al.},
  booktitle={2019 IEEE/CVF International Conference on Computer Vision Workshop (ICCVW)}, 
  title={Image Super-Resolution via Attention Based Back Projection Networks}, 
  year={2019},

  pages={3517-3525},
}

@ARTICLE{sr_2,
  author={Liu, Zhi-Song and et al.},
  journal={IEEE Transactions on Circuits and Systems for Video Technology}, 
  title={Photo-Realistic Image Super-Resolution via Variational Autoencoders}, 
  year={2021},
  volume={31},
  number={4},
  pages={1351-1365},
}

@InProceedings{sr_4,
author = {Liu, Zhi-Song and et al.},
title = {Unsupervised Real Image Super-Resolution via Generative Variational AutoEncoder},
booktitle = {Proceedings of the IEEE/CVF Conference on Computer Vision and Pattern Recognition (CVPR) Workshops},
month = {June},
year = {2020}
}

@inproceedings{han,
author = {Niu, Ben and Wen, Weilei and Ren, Wenqi and Zhang, Xiangde and Yang, Lianping and Wang, Shuzhen and Zhang, Kaihao and Cao, Xiaochun and Shen, Haifeng},
title = {{Single Image Super-Resolution via a Holistic Attention Network}},
year = {2020},
isbn = {978-3-030-58609-6},
booktitle = {European Confernece on Computer Vision (ECCV2020)},
pages = {191–207},
numpages = {17},
location = {Glasgow, United Kingdom}
}

@InProceedings{hat,
    author    = {Chen, Xiangyu and Wang, Xintao and Zhou, Jiantao and Qiao, Yu and Dong, Chao},
    title     = {{Activating More Pixels in Image Super-Resolution Transformer}},
    booktitle = {Proceedings of the IEEE/CVF Conference on Computer Vision and Pattern Recognition (CVPR)},
    month     = {June},
    year      = {2023},
    pages     = {22367-22377}
}

@InProceedings{drct,
  author    = {Hsu, Chih-Chung and Lee, Chia-Ming and Chou, Yi-Shiuan},
  title     = {{DRCT: Saving Image Super-Resolution Away from Information Bottleneck}},
  booktitle = {Proceedings of the IEEE/CVF Conference on Computer Vision and Pattern Recognition (CVPR) Workshops},
  month     = {June},
  year      = {2024},
  pages     = {6133-6142}
}

@inproceedings{swin,
  title={{Swin Transformer: Hierarchical Vision Transformer using Shifted Windows}},
  author={Liu, Ze and Lin, Yutong and Cao, Yue and Hu, Han and Wei, Yixuan and Zhang, Zheng and Lin, Stephen and Guo, Baining},
  booktitle={Proceedings of the IEEE/CVF International Conference on Computer Vision (ICCV)},
  year={2021}
}

@InProceedings{resshift,
  author    = {Zongsheng Yue and Jianyi Wang and Chen Change Loy},
  title     = {{ResShift: Efficient Diffusion Model for Image Super-resolution by Residual Shifting}},
  booktitle = {Advances in Neural Information Processing Systems (NeurIPS)},
  year      = {2023},
}

@InProceedings{attention,
      title={Attention is all you need}, 
      author={A. Vaswani and N. Shazeer and N. Parmar and J. Uszkoreit and L. Jones and A. Gomez, I. Polosukhin},
      year={2017},
      booktitle= {Advances in neural information processing systems},
      pages= {5998-6008}
}

@inproceedings{lam,
  title={{Interpreting Super-Resolution Networks with Local Attribution Maps}},
  author={Gu, Jinjin and Dong, Chao},
  booktitle={Proceedings of the IEEE/CVF Conference on Computer Vision and Pattern Recognition},
  pages={9199--9208},
  year={2021}
}

@InProceedings{srgan,
  author={Ledig, Christian and Theis, Lucas and Huszár, Ferenc and Caballero, Jose and Cunningham, Andrew and Acosta, Alejandro and Aitken, Andrew and Tejani, Alykhan and Totz, Johannes and Wang, Zehan and Shi, Wenzhe},
  booktitle={2017 IEEE Conference on Computer Vision and Pattern Recognition (CVPR)}, 
  title={{Photo-Realistic Single Image Super-Resolution Using a Generative Adversarial Network}}, 
  year={2017},
  volume={},
  number={},
  pages={105-114},
  doi={10.1109/CVPR.2017.19}
}

@InProceedings{lte,
    author    = {Lee, Jaewon and Jin, Kyong Hwan},
    title     = {{Local Texture Estimator for Implicit Representation Function}},
    booktitle = {Proceedings of the IEEE/CVF Conference on Computer Vision and Pattern Recognition (CVPR)},
    month     = {June},
    year      = {2022},
    pages     = {1929-1938}
}

@InProceedings{esrgan,
    author = {Wang, Xintao and Yu, Ke and Wu, Shixiang and Gu, Jinjin and Liu, Yihao and Dong, Chao and Qiao, Yu and Loy, Chen Change},
    title = {{ESRGAN: Enhanced super-resolution generative adversarial networks}},
    booktitle = {The European Conference on Computer Vision Workshops (ECCVW)},
    month = {September},
    year = {2018}
}

@inproceedings{gan,
author = {Goodfellow, Ian J. and Pouget-Abadie, Jean and Mirza, Mehdi and Xu, Bing and Warde-Farley, David and Ozair, Sherjil and Courville, Aaron and Bengio, Yoshua},
title = {Generative adversarial nets},
year = {2014},
booktitle = { International Conference on Neural Information Processing Systems},
pages = {2672–2680},
numpages = {9},
location = {Montreal, Canada},
series = {NIPS'14}
}

@InProceedings{ddpm,
  title = 	 {{Deep Unsupervised Learning using Nonequilibrium Thermodynamics}},
  author = 	 {Sohl-Dickstein, Jascha and Weiss, Eric and Maheswaranathan, Niru and Ganguli, Surya},
  booktitle = 	 {Proceedings of the 32nd International Conference on Machine Learning},
  pages = 	 {2256--2265},
  year = 	 {2015},
  editor = 	 {Bach, Francis and Blei, David},
  volume = 	 {37},
  series = 	 {Proceedings of Machine Learning Research},
  address = 	 {Lille, France},
  month = 	 {07--09 Jul},
  publisher =    {PMLR},
}

@inproceedings{scorebased,
  title={{Score-Based Generative Modeling through Stochastic Differential Equations}},
  author={Yang Song and Jascha Sohl-Dickstein and Diederik P Kingma and Abhishek Kumar and Stefano Ermon and Ben Poole},
  booktitle={International Conference on Learning Representations},
  year={2021},
  url={https://openreview.net/forum?id=PxTIG12RRHS}
}

@inproceedings{cold,
title={{Cold Diffusion: Inverting Arbitrary Image Transforms Without Noise}},
author={Arpit Bansal and Eitan Borgnia and Hong-Min Chu and Jie S. Li and Hamid Kazemi and Furong Huang and Micah Goldblum and Jonas Geiping and Tom Goldstein},
booktitle={Thirty-seventh Conference on Neural Information Processing Systems},
year={2023},
url={https://openreview.net/forum?id=XH3ArccntI}
}

@inproceedings{autoregressive,
title={{Autoregressive Diffusion Models}},
author={Emiel Hoogeboom and Alexey A. Gritsenko and Jasmijn Bastings and Ben Poole and Rianne van den Berg and Tim Salimans},
booktitle={International Conference on Learning Representations},
year={2022},
url={https://openreview.net/forum?id=Lm8T39vLDTE}
}

@INPROCEEDINGS{ldm,
  author={Rombach, Robin and Blattmann, Andreas and Lorenz, Dominik and Esser, Patrick and Ommer, Björn},
  booktitle={2022 IEEE/CVF Conference on Computer Vision and Pattern Recognition (CVPR)}, 
  title={{High-Resolution Image Synthesis with Latent Diffusion Models}}, 
  year={2022},
  volume={},
  number={},
  pages={10674-10685},
  doi={10.1109/CVPR52688.2022.01042}}

@inproceedings{ide,
title={{Idempotent Generative Network}},
author={Assaf Shocher and Amil V Dravid and Yossi Gandelsman and Inbar Mosseri and Michael Rubinstein and Alexei A Efros},
booktitle={The Twelfth International Conference on Learning Representations},
year={2024},
url={https://openreview.net/forum?id=XIaS66XkNA}
}

@inproceedings{consistency,
author = {Song, Yang and others},
title = {{Consistency Models}},
year = {2023},
booktitle = {Proceding of International Conference on Machine Learning},
articleno = {1335},
numpages = {42},
location = {Honolulu, Hawaii, USA},
series = {ICML'23}
}

@article{sr3,
  author={Saharia, Chitwan and Ho, Jonathan and Chan, William and Salimans, Tim and Fleet, David J. and Norouzi, Mohammad},
  journal={IEEE Transactions on Pattern Analysis and Machine Intelligence}, 
  title={{Image Super-Resolution via Iterative Refinement}}, 
  year={2023},
  volume={45},
  number={4},
  pages={4713-4726},
  doi={10.1109/TPAMI.2022.3204461}}

@article{toflow,
  title={Video Enhancement with Task-Oriented Flow},
  author={Xue, Tianfan and Chen, Baian and Wu, Jiajun and Wei, Donglai and Freeman, William T},
  journal={{International Journal of Computer Vision (IJCV)}},
  volume={127},
  number={8},
  pages={1106--1125},
  year={2019},
  publisher={Springer}
}

@INPROCEEDINGS{duf,
  author={Jo, Younghyun and Oh, Seoung Wug and Kang, Jaeyeon and Kim, Seon Joo},
  booktitle={2018 IEEE/CVF Conference on Computer Vision and Pattern Recognition}, 
  title={{Deep Video Super-Resolution Network Using Dynamic Upsampling Filters Without Explicit Motion Compensation}}, 
  year={2018},
  volume={},
  number={},
  pages={3224-3232}}

@INPROCEEDINGS{tdan,
  author={Tian, Yapeng and Zhang, Yulun and Fu, Yun and Xu, Chenliang},
  booktitle={2020 IEEE/CVF Conference on Computer Vision and Pattern Recognition (CVPR)}, 
  title={{TDAN: Temporally-Deformable Alignment Network for Video Super-Resolution}}, 
  year={2020},
  volume={},
  number={},
  pages={3357-3366},
  doi={10.1109/CVPR42600.2020.00342}}

@InProceedings{upscale,
    author    = {Zhou, Shangchen and Yang, Peiqing and Wang, Jianyi and Luo, Yihang and Loy, Chen Change},
    title     = {{Upscale-A-Video: Temporal-Consistent Diffusion Model for Real-World Video Super-Resolution}},
    booktitle = {Proceedings of the IEEE/CVF Conference on Computer Vision and Pattern Recognition (CVPR)},
    month     = {June},
    year      = {2024},
    pages     = {2535-2545}
}

@ARTICLE{vrt,
  author={Liang, Jingyun and Cao, Jiezhang and Fan, Yuchen and Zhang, Kai and Ranjan, Rakesh and Li, Yawei and Timofte, Radu and Van Gool, Luc},
  journal={IEEE Transactions on Image Processing}, 
  title={{VRT: A Video Restoration Transformer}}, 
  year={2024},
  volume={33},
  number={},
  pages={2171-2182},
  doi={10.1109/TIP.2024.3372454}}

@inproceedings{MeshfreeFlowNet,
author = {Jiang, Chiyu "Max" and Esmaeilzadeh, Soheil and Azizzadenesheli, Kamyar and Kashinath, Karthik and Mustafa, Mustafa and Tchelepi, Hamdi A. and Marcus, Philip and Prabhat and Anandkumar, Anima},
title = {{MeshfreeFlowNet: a physics-constrained deep continuous space-time super-resolution framework}},
year = {2020},
isbn = {9781728199986},
booktitle = {Proceedings of the International Conference for High Performance Computing, Networking, Storage and Analysis},
articleno = {9},
numpages = {15},
location = {Atlanta, Georgia},
series = {SC '20}
}

@article{piesrgan,
title = {Using physics-informed enhanced super-resolution generative adversarial networks for subfilter modeling in turbulent reactive flows},
journal = {Proceedings of the Combustion Institute},
volume = {38},
number = {2},
pages = {2617-2625},
year = {2021},
issn = {1540-7489},
doi = {https://doi.org/10.1016/j.proci.2020.06.022},
author = {Mathis Bode and Michael Gauding and Zeyu Lian and Dominik Denker and Marco Davidovic and Konstantin Kleinheinz and Jenia Jitsev and Heinz Pitsch}
}

@article{Fukami,
title = {Super-resolution reconstruction of turbulent flows with machine learning},
journal = {Journal of Fluid Mechanics},
volume = {870},
pages = {106-120},
year = {2019},
doi = {10.1017/jfm.2019.238},
author = {K. Fukami and K. Fukagata and and K. Taira}
}

@misc{harder_1,
author = {Harder, Paula and Yang, Qidong and et al.},
  title = {Generating physically-consistent high-resolution climate data with hard-constrained neural networks},
  publisher = {arXiv}, 
  year = {2022}
}

@article{beucler,
      title={{Achieving Conservation of Energy in Neural Network Emulators for Climate Modeling}}, 
      author={Tom Beucler and Stephan Rasp and Michael Pritchard and Pierre Gentine},
      year={2019},
      eprint={1906.06622},
      archivePrefix={arXiv},
      primaryClass={physics.ao-ph},
      url={https://arxiv.org/abs/1906.06622}, 
}

@article{harder_2, 
    title={Physics-informed learning of aerosol microphysics}, 
    volume={1}, 
    DOI={10.1017/eds.2022.22}, 
    journal={Environmental Data Science}, 
    author={Harder, Paula and Watson-Parris, Duncan and et al.}, 
    year={2022}, 
    pages={e20}
}

@inproceedings{lstm,
author = {Shi, Xingjian and Chen, Zhourong and Wang, Hao and Yeung, Dit-Yan and Wong, Wai-kin and Woo, Wang-chun},
title = {{Convolutional LSTM Network: a machine learning approach for precipitation nowcasting}},
year = {2015},
address = {Cambridge, MA, USA},
booktitle = {Proceedings of the 28th International Conference on Neural Information Processing Systems},
pages = {802–810},
numpages = {9},
location = {Montreal, Canada},
series = {NIPS'15}
}

@article{alphaflow,
  author		= "Amanda A. Volk and Robert W. Epps and Daniel T. Yonemoto and Benjamin S. Masters et al.",
  title		= "{AlphaFlow}: autonomous discovery and optimization of multi-step chemistry using a self-driven fluidic lab guided by reinforcement learning",
  journal		= "Nature Communication",
  number		= "1403",
  year			= "2023",
}

@article{material2,
  author		= "Goodall, R. E. A. and Lee, A. A.",
  title		= "Predicting materials properties without crystal structure: deep representation learning from stoichiometry",
  journal		= "Nature Communication",
  number		= "11",
  volume={6280},
  year			= "2020",
}

@article{graphcast,
  author		= "Remi Lam et al.",
  title		= "Learning skillful medium-range global weather forecasting",
  journal		= "Science",
  pages		= "1416--1421",
  volume = {382},
  year			= "2023",
}

@article{pangu,
      title={Accurate medium-range global weather forecasting with {{3D}} neural networks}, 
      author={Bi, K. and Xie, L. and Zhang, H. et al.},
      journal		= "Nature",
      pages		= "533--538",
      volume = {619},
      year			= "2023",
}

@article{smoke,
author = {Bai, Kai and Li, Wei and Desbrun, Mathieu and Liu, Xiaopei},
title = {{Dynamic Upsampling of Smoke through Dictionary-based Learning}},
year = {2020},
issue_date = {February 2021},
publisher = {Association for Computing Machinery},
address = {New York, NY, USA},
volume = {40},
number = {1},
issn = {0730-0301},
doi = {10.1145/3412360},
journal = {ACM Trans. Graph.},
month = {sep},
articleno = {4},
numpages = {19},
}

@INPROCEEDINGS{physrnet,
  author={Arora, Rajat},
  booktitle={2022 IEEE/ACM International Workshop on Artificial Intelligence and Machine Learning for Scientific Applications (AI4S)}, 
  title={{PhySRNet: Physics informed super-resolution network for application in computational solid mechanics}}, 
  year={2022},
  volume={},
  number={},
  pages={13-18},
  doi={10.1109/AI4S56813.2022.00008}}

@article{gao,
    author = {Gao, Han and Sun, Luning and Wang, Jian-Xun},
    title = {Super-resolution and denoising of fluid flow using physics-informed convolutional neural networks without high-resolution labels},
    journal = {Physics of Fluids},
    volume = {33},
    number = {7},
    pages = {073603},
    year = {2021},
    month = {07},
    issn = {1070-6631},
    doi = {10.1063/5.0054312},
}

@article{Teufel,
    author = {Teufel B. and Carmo, F. and Sushama, L. et al.},
    title = {Physics-informed deep learning framework to model intense precipitation events at super resolution},
    journal = {Geoscience Letter},
    volume = {10},
    number = {19},
    year = {2023},
    month = {07},
    doi = {10.1186/s40562-023-00272-z},
}

@article{shu,
title = {A physics-informed diffusion model for high-fidelity flow field reconstruction},
journal = {Journal of Computational Physics},
volume = {478},
pages = {111972},
year = {2023},
issn = {0021-9991},
doi = {https://doi.org/10.1016/j.jcp.2023.111972},
author = {Dule Shu and Zijie Li and Amir {Barati Farimani}},
}

@article{Gerhard2015,
title = {{Multiwavelet-based grid adaptation with discontinuous Galerkin schemes for shallow water equations}},
journal = {Journal of Computational Physics},
volume = {301},
pages = {265-288},
year = {2015},
issn = {0021-9991},
doi = {https://doi.org/10.1016/j.jcp.2015.08.030},
author = {Nils Gerhard and Daniel Caviedes-Voullième and Siegfried Müller and Georges Kesserwani},
}

@Article{sst,
AUTHOR = {Fanelli, C. and Ciani, D. and Pisano, A. and Buongiorno Nardelli, B.},
TITLE = {Deep learning for the super resolution of {M}editerranean sea surface temperature fields},
JOURNAL = {Ocean Science},
VOLUME = {20},
YEAR = {2024},
NUMBER = {4},
PAGES = {1035--1050},
URL = {https://os.copernicus.org/articles/20/1035/2024/},
DOI = {10.5194/os-20-1035-2024}
}

@article{sst_2,
title = {{Multi-source deep data fusion and super-resolution for downscaling sea surface temperature guided by Generative Adversarial Network-based spatiotemporal dependency learning}},
journal = {International Journal of Applied Earth Observation and Geoinformation},
volume = {119},
pages = {103312},
year = {2023},
issn = {1569-8432},
doi = {https://doi.org/10.1016/j.jag.2023.103312},
author = {Jinah Kim and Taekyung Kim and Joon-Gyu Ryu},
}

@article{ssh,
title = {{Downscaling of ocean fields by fusion of heterogeneous observations using Deep Learning algorithms}},
journal = {Ocean Modelling},
volume = {182},
pages = {102174},
year = {2023},
issn = {1463-5003},
doi = {https://doi.org/10.1016/j.ocemod.2023.102174},
author = {Sylvie Thiria and Charles Sorror and Theo Archambault and Anastase Charantonis and Dominique Bereziat and Carlos Mejia and Jean-Marc Molines and Michel Crépon},
}

@article{ssh_2,
title = {Downscaling sea surface height and currents in coastal regions using convolutional neural network},
journal = {Applied Ocean Research},
volume = {151},
pages = {104153},
year = {2024},
issn = {0141-1187},
doi = {https://doi.org/10.1016/j.apor.2024.104153},
author = {Bing Yuan and Benjamin Jacob and Wei Chen and Joanna Staneva},
}

@article{gwm,
title = {Accurate medium-range global weather forecasting with {3D} neural networks},
journal = {Nature},
volume = {619},
pages = {533-538},
year = {2023},
doi = {https://doi.org/10.1038/s41586-023-06185-3},
author = {Bi, K. and Xie, L. and Zhang, H. et al.},
}

@article{climax,
  title={{ClimaX: A foundation model for weather and climate}},
  author={Nguyen, Tung and Brandstetter, Johannes and Kapoor, Ashish and Gupta, Jayesh K and Grover, Aditya},
  journal={Conference on Neural Information Processing Systems (NeurIPS)},
  year={2023}
}

@article{kubatko2006hp,
  title={hp discontinuous {Galerkin} methods for advection dominated problems in shallow water flow},
  author={Kubatko, Ethan J and Westerink, Joannes J and Dawson, Clint},
  journal={Computer Methods in Applied Mechanics and Engineering},
  volume={196},
  number={1-3},
  pages={437--451},
  year={2006},
  publisher={Elsevier}
}

@article{contreras2023channel,
  title={{A channel-to-basin scale ADCIRC based hydrodynamic unstructured mesh model for the US East and Gulf of Mexico coasts}},
  author={Contreras, Maria Teresa and Woods, Brendan and Blakely, Coleman and Wirasaet, Damrongsak and Westerink, Joannes and Cobell, Zach and Pringle, William and Moghimi, Saeed and Vinogradov, Sergey and Myers, Edward and Seroka, Greg and Lalime, Michael and Funakoshi, Yuji and Van der Westhuysen, Andre and Abdolali, Ali and Valseth, Eirik and Dawson, Clint;},
  year={2023},
  doi = {https://doi.org/10.25923/wktm-c719},
  publisher = {NOAA},
  journal = { NOAA technical memorandum NOS CS 54}
}

@article{brown2010atlantic,
  title={Atlantic hurricane season of 2008},
  author={Brown, Daniel P and Beven, John L and Franklin, James L and Blake, Eric S},
  journal={Monthly Weather Review},
  volume={138},
  number={5},
  pages={1975--2001},
  year={2010}
}

@article{hope2013hindcast,
  title={Hindcast and validation of Hurricane Ike (2008) waves, forerunner, and storm surge},
  author={Hope, Mark E and Westerink, Joannes J and Kennedy, Andrew B and Kerr, PC and Dietrich, J Casey and Dawson, C and Bender, Christopher J and Smith, JM and Jensen, Robert E and Zijlema, Marcel and others},
  journal={Journal of Geophysical Research: Oceans},
  volume={118},
  number={9},
  pages={4424--4460},
  year={2013},
  publisher={Wiley Online Library}
}

@article{bunya2009wetting,
  title={{A wetting and drying treatment for the Runge--Kutta discontinuous Galerkin solution to the shallow water equations}},
  author={Bunya, Shintaro and Kubatko, Ethan J and Westerink, Joannes J and Dawson, Clint},
  journal={Computer Methods in Applied Mechanics and Engineering},
  volume={198},
  number={17-20},
  pages={1548--1562},
  year={2009},
  publisher={Elsevier}
}

@article{egbert2002efficient,
  title={Efficient inverse modeling of barotropic ocean tides},
  author={Egbert, Gary D and Erofeeva, Svetlana Y},
  journal={Journal of Atmospheric and Oceanic technology},
  volume={19},
  number={2},
  pages={183--204},
  year={2002}
}

@article {Westerink1989,
      author = "J. J.  Westerink and K. D.  Stolzenbach and J. J.  Connor",
      title = "General Spectral Computations of the Nonlinear Shallow Water Tidal Interactions within the Bight of Abaco",
      journal = "Journal of Physical Oceanography",
      year = "1989",
      publisher = "American Meteorological Society",
      address = "Boston MA, USA",
      volume = "19",
      number = "9",
      doi = "10.1175/1520-0485(1989)019<1348:GSCOTN>2.0.CO;2",
      pages=      "1348 - 1371",
      url = "https://journals.ametsoc.org/view/journals/phoc/19/9/1520-0485_1989_019_1348_gscotn_2_0_co_2.xml"
}

@article{Gottlieb2001,
author = {Gottlieb, Sigal and Shu, Chi-Wang and Tadmor, Eitan},
title = {Strong Stability-Preserving High-Order Time Discretization Methods},
journal = {SIAM Review},
volume = {43},
number = {1},
pages = {89-112},
year = {2001},
doi = {10.1137/S003614450036757X},
URL = { https://doi.org/10.1137/S003614450036757X},
eprint = { https://doi.org/10.1137/S003614450036757X}
}

@TechReport{adcirc,
  author      = {Luettich, R. and Westerink, J. and Scheffner, N.},
  institution = {US Army Engineers Waterways Experiment Station, Vicksburg, MS},
  title       = {{ADCIRC: A}n advanced three-dimensional circulation model for shelves, coasts and estuaries, {R}eport~1: {T}heory and methodology of {ADCIRC-2DDI} and {ADCIRC-3DL}},
  year        = {1992},
  number      = {Dredging Research Program Technical Report DRP-92-6},
}

@ARTICLE{Kernkamp2011,
	author = {Kernkamp, Herman W. J. and Van Dam, Arthur and Stelling, Guus S. and De Goede, Erik D.},
	title = {Efficient scheme for the shallow water equations on unstructured grids with application to the {C}ontinental {S}helf},
	year = {2011},
	journal = {Ocean Dynamics},
	volume = {61},
	number = {8},
	pages = {1175 – 1188},
	doi = {10.1007/s10236-011-0423-6},
	url = {https://www.scopus.com/inward/record.uri?eid=2-s2.0-81255137954&doi=10.1007%2fs10236-011-0423-6&partnerID=40&md5=bccab7c0af7a6e52ebbc1c1e37351443},
	type = {Article}
}

@article{DuebenKA2012,
Author = {P. Düben and P. Korn and V. Aizinger},
Title = {A discontinuous/continuous low order finite element shallow water model on the sphere},
Journal = {Journal of Computational Physics},
Year = {2012},
Volume = {231},
Number = {6},
Pages = {2396-2413},
Publisher = {Elsevier},
Doi = {10.1016/j.jcp.2011.11.018},
Url = {www.sciencedirect.com/science/article/pii/S0021999111006735}
}

@misc{wichitrnithed2025,
      title={{GPU-acceleration of the Discontinuous Galerkin Shallow Water Equations Model (DG-SWEM) with OpenACC}}, 
      author={Chayanon Wichitrnithed and Eirik Valseth and Ethan J. Kubatko and Shintaro Bunya and Clint Dawson},
      year={2025},
      eprint={2508.21208},
      archivePrefix={arXiv},
      primaryClass={physics.comp-ph},
      url={https://arxiv.org/abs/2508.21208}, 
}

\end{document}